\documentclass[preprint,12pt]{elsarticle}
\usepackage[boxed,noline]{algorithm2e}
\usepackage{amssymb}
\usepackage{graphicx}
\usepackage{multirow}
\usepackage{setspace}
\usepackage{subcaption}
\usepackage{threeparttable}
\usepackage{hyperref}
\journal{Computer Networks}

\begin{document}

\begin{frontmatter}

  \title{The Effect of ISP Traffic Shaping on User-Perceived Performance in
    Broadband Shared Access Networks}

  \author[kk]{Kyeong Soo Kim\corref{cor}}%%
  \cortext[cor]{Tel.: +44 (0)1792 602024.}%%
  \ead{k.s.kim@swansea.ac.uk}%%
  \address[kk]{College of Engineering, Swansea University, Swansea, SA2 8PP,
    Wales United Kingdom}%%

  \begin{abstract}
    Recent studies on the practice of shaping subscribers' traffic by Internet
    service providers (ISPs) give a new insight into the actual performance of
    broadband access networks at a packet level. Unlike metro and backbone
    networks, however, access networks directly interface with end-users, so it
    is important to base the study and design of access networks on the
    behaviors of and the actual performance perceived by end-users. In this
    paper we study the effect of ISP traffic shaping using traffic models based
    on user behaviors and application/session-layer metrics providing
    quantifiable measures of user-perceived performance for HTTP, FTP, and
    streaming video traffic. To compare the user-perceived performance of shaped
    traffic flows with those of unshaped ones in an integrated way, we use a
    multivariate non-inferiority testing procedure. We first investigate the
    effect of the token generation rate and the token bucket size of a token
    bucket filter (TBF) on user-perceived performance at a subscriber level with
    a single subscriber. Then we investigate their effect at an access level
    where shaped traffic flows from multiple subscribers interact with one
    another in a common shared access network. The simulation results show that
    for a given token generation rate, a larger token bucket --- i.e., up to 100
    MB and 1 GB for access line rates of 100 Mbit/s and 1 Gbit/s, respectively
    --- provides better user-perceived performance at both subscriber and access
    levels. It is also shown that the loose burst control resulting from the
    large token bucket --- again up to 100 MB for access line rate of 100 Mbit/s
    --- does not negatively affect user-perceived performance with multiple
    subscribers even in the presence of non-conformant subscribers; with a much
    larger token bucket (e.g., size of 10 GB), however, the negative effect of
    non-conformant subscribers on the user-perceived performance of conformant
    subscribers becomes clearly visible because the impact of token bucket size
    and that of token generation rate are virtually indistinguishable in this
    case.
  \end{abstract}


\end{frontmatter}

\section{Introduction}
\label{sec-1}
The practice of shaping subscribers' traffic by Internet service providers
(ISPs) has been under intensive study; for example, the effect of ISP traffic
shaping on various packet-level performance with its detection and policies has
been investigated based on actual measurements and mathematical/simulation
analyses in \cite{bauer11:_power}, \cite{sundaresan11:_broad_inter_perfor},
\cite{kanuparthy11:_end_isp}, and \cite{marcon11}, which provides a new insight
into the actual performance of broadband access networks at a packet level.

Unlike metro and backbone networks, however, access networks directly interface
with end-users, so it is important to base the study and design of access
networks on the behaviors of and the actual performance perceived by end-users
\cite{Kim:11-4}. The major goal of our study in this paper, therefore, is to
investigate the effect of ISP traffic shaping on user-perceived performance,
i.e., the quality of experience (QoE), and thereby to provide ISPs further
insights into the design, deployment, and operation of the next-generation
access networks from end-users' perspective. Because in access networks the
average rate of a service is determined by a service contract between a
subscriber and an ISP (e.g., through subscription tiers), which is then
controlled accordingly by the token generation rate of a token bucket filter
(TBF) used in traffic shaping, and the peak rate is usually determined by the
underlying access technology (e.g., line rates of digital subscriber line (DSL)
and cable Internet), we put our major focus on the effect of the token bucket
size of a TBF on user-perceived performance given the token generation rate and
the peak rate of an access link.\footnote{The token generation rate and the
  token bucket size correspond to the maximum sustained traffic rate (MSTR) and
  the maximum traffic burst in the data over cable service interface
  specifications (DOCSIS) media access control (MAC) and upper layer protocols
  interface specification \cite{sp-mulpiv3.0}, respectively. }

The current study was specifically triggered by the results from recent
investigations of ``PowerBoost'' in \cite{bauer11:_power} and
\cite{sundaresan11:_broad_inter_perfor}, the feature present in some cable
broadband networks that enables sharing of unused capacity by giving customers
extra bursts of speed whose duration is controlled by the token bucket size. In
this paper we extend those investigations of the effect of the token bucket size
on packet-level performance at network/transport layers to those on
user-perceived performance at application/session layers based on the research
framework which we proposed for the clean-slate design of next-generation access
networks \cite{Kim:11-2,Kim:11-4}.

Through the investigation we answer the following key questions:
\begin{itemize}
\item \textbf{Subscriber level}: If we consider a single subscriber under
  traffic shaping in isolation, what is the minimum token bucket size providing
  user-perceived performance \emph{nearly equivalent}\footnote{The term
    ``nearly equivalent'' is formally defined based on multivariate
    non-inferiority testing in Sec. \ref{sec-3-3}.} to those of a subscriber
  under no traffic shaping for a given token generation rate and a mix of
  traffic flows?
\end{itemize}
\begin{itemize}
\item \textbf{Access level}: In a shared access, where shaped traffic flows from
  multiple subscribers interact with one another, what is the effect of the
  token bucket size on user-perceived performance of all subscribers?
  Specifically, how many subscribers can be served with user-perceived
  performance nearly equivalent to those of a subscriber under no traffic
  shaping in a dedicated access for a given token generation rate and a mix of
  traffic flows as well as a token bucket size suggested by the subscriber-level
  investigation?
\end{itemize}
It is this \emph{trade-off in traffic shaping} between the performance at
subscriber and access levels that interests both subscribers and
ISPs. Considering the bursty nature of traffic flows at multiple layers, e.g.,
user behaviors and variable bit rate (VBR) encoding at the application/session
layers and transmission control protocol (TCP) flow and congestion controls at
the transport layer, one can expect that at the subscriber level, the
user-perceived performance of a subscriber under traffic shaping with loose
burst control would approach those of a subscriber under no traffic shaping when
the token generation rate is equal to or greater than the long-term average rate
of combined traffic flows. At the access level, on the other hand, it is likely
that the loose burst control at the subscriber level negatively affects the
performance of other subscribers, especially when there are non-conformant or
mis-behaving subscribers.\footnote{Subscribers who consistently generate traffic
  whose long-term average is higher than that of the service contract (i.e., the
  token generation rate) are called \emph{non-conformant} or \emph{mis-behaving}
  subscribers in this paper. }

Note that the traffic shaping and related issues (e.g., multiplexing and
scheduling of shaped flows) have been extensively studied in the context of
per-flow/connection traffic shaping and based on packet-level measures since the
introduction of the ``leaky bucket'' method in \cite{Turner:86}.\footnote{The
  leaky bucket algorithm described in \cite{Turner:86} is basically the same as
  the token bucket algorithm. We use the terms ``leaky bucket'' and ``token
  bucket'' interchangeably in this paper. } In
\cite{Elwalid:91,Veciana:94,Kim:92,Yin:91}, the TBF and their analyses with
various statistical traffic models are studied. In
\cite{garroppo02:_estim_voip,procissi02:_token}, the dimensioning of TBF
parameters through the notion of a linear-bounded arrival process (LBAP) is
investigated for aggregated voice over Internet protocol (VoIP) and long-range
dependent (LRD) traffic.  In \cite{Chong:96}, the performance trade-off of
traffic shaping between access control queueing and network queueing is studied
based on the spectral analysis technique, while in \cite{bregni10:_charac}, the
characterization of LRD traffic regulated by leaky-bucket policers and shapers
is studied using the modified Allan variance (MAVAR) for the LRD estimation and
spectral analysis of the regulated traffic. As for scheduling of shaped traffic
flows, the end-to-end delay bounds and the buffer space requirements of various
scheduling disciplines are well summarized in \cite{zhang95:_servic}.

The results from these studies suggest that given a token generation rate,
allowing large bursts through a large token bucket size improves the
packet-level performance of an individual flow, while multiplexing of those
shaped flows would increase the deterministic bound of end-to-end packet
delay. These works, however, are not done in the context of ISP traffic shaping,
where multiple traffic flows with different service types are shaped together by
a single TBF, and do not take into account user behaviors in traffic generation
and performance perceived by end-users. To the best of our knowledge, our work
in this paper is the first attempt to systematically assess the effect of ISP
traffic shaping on user-perceived performance with user-behavior-based traffic
models at both subscriber and access levels.

The rest of the paper is organized as follows: Section \ref{sec-2} provides an
overview of the current practice of ISP traffic shaping and its major
issues. Section \ref{sec-3} describes the methodology we adopt for this
investigation with details of experimental setup and a comparative analysis
framework. Section \ref{sec-4} presents the results of experimental
investigation of the effect of ISP traffic shaping at both subscriber and access
levels. Section \ref{sec-5} concludes our work in this paper.

\section{Overview of ISP Traffic Shaping}
\label{sec-2}

\subsection{Current Practice}
\label{sec-2-1}
Traffic shaping was originally devised for connection-oriented networks to
regulate an \emph{individual flow} per traffic conformance definition negotiated
during a connection admission control (CAC) at a user-network interface (UNI)
\cite{Turner:86,I.371:93}. It is now used by ISPs to regulate \emph{combined
  flows} from a subscriber in a different context of connectionless IP networks:
Because there are no CAC procedures used at the UNI in the current IP-based
networks, ISPs base their traffic shaping on service contracts with subscribers,
which are informal compared to standard traffic conformance definitions (e.g.,
those for guaranteed quality of service (QoS) in Internet \cite{rfc:2212}).

Typically ISPs use traffic shaping to divide the available capacity of a
physical access link (e.g., 100+ Mbps by DOCSIS 3.0 \cite{sp-mulpiv3.0}) into
smaller ones promised to their subscribers per service contracts
\cite{lakshminarayanan04:_bandw}. With the mechanism like TBF, ISPs regulate the
token generation rate, the token bucket size, and optionally the peak rate of
combined traffic flows from each subscriber, which provides reasonable QoS to
conformant subscribers but prevents non-conformant subscribers from hogging the
available bandwidth. At the same time, ISPs want to allow efficient sharing of
unused capacity among active subscribers to improve their experience of Internet
access, which, like the PowerBoost, is a way to differentiate their access
services from their competitors \cite{bauer11:_power}.

\subsection{Major Issues}
\label{sec-2-2}
The current practice of ISP traffic shaping incurs the following major issues
due to its application to the combined traffic flows from a subscriber and the
lack of formal definition of traffic conformance as we discussed.

\subsubsection{Service Differentiation}
\label{sec-2-2-1}
Under the current practice of ISP traffic shaping, it is difficult to provide
different levels of QoS to different types of traffic flows. For instance, when
there are delay-sensitive flows (e.g., VoIP calls) and large non-real-time data
flows (e.g., file transfer) from the same subscriber, the current ISP traffic
shaping cannot differentiate the former from the latter because it is done per
subscriber over combined flows. If traffic shaping is done per individual flow
as in integrated services (IntServ) \cite{rfc:1633} or at least per class as in
differentiated services (DiffServ) \cite{rfc:2475}, this issue can be
addressed. To do that, however, we need per-flow or per-class service contracts,
which is not the case currently.

As workarounds, two traffic control schemes for large bulk data flows under the
PowerBoost --- one based on intermittent transmission with periodic \emph{on}
and \emph{off} cycles and the other based on WonderShaper --- are investigated
in \cite{sundaresan11:_broad_inter_perfor}, which significantly improve the
latency of delay-sensitive flows while achieving similar long-term rates because
these schemes do not deplete the tokens at any time and thereby remove the
chance of queueing. The main drawback is that the shaping parameters need to be
known in order to exploit this behavior.

\subsubsection{Conflicting Requirements}
\label{sec-2-2-2}
Another major issue is that guaranteeing QoS to the subscribers and enabling
efficient sharing of unused capacity among them seemingly contradict each
other. For better QoS guarantee, tight burst control is preferred for stricter
control of user traffic; for efficient sharing of unused capacity among active
subscribers, on the other hand, loose burst control is desirable as in the
PowerBoost. Because the average and the peak rates of a service are determined
by a service contract and underlying access technology respectively, determining
a proper size of the token bucket is a key to ISP traffic shaping.

The effect of ISP traffic shaping, especially the effect of the token bucket
size, has been studied with the PowerBoost: In \cite{bauer11:_power}, a
qualitative investigation of the effect of PowerBoost on TCP and applications is
done, while its impact on ISP speed measurements is studied based on the actual
results from SamKnows measurements. In \cite{sundaresan11:_broad_inter_perfor},
the effect of the PowerBoost is also studied based on the results obtained from
two independent gateway deployments with focus on packet-level performance at
the network/transport layers like packet latency and TCP throughput. Even though
both studies are based on the measurements from field-deployed home gateways and
thereby provide meaningful snapshots of actual broadband access performance,
there is neither systematic investigation on the effect of ISP traffic shaping
on user-perceived performance nor consistent conclusion made even for
packet-level performance because so many conditions, including the way and the
time of measurements and background traffic from other users, are simply out of
control in such large-scale field tests. They also lack the investigation of the
interaction of shaped traffic flows from multiple subscribers in a common shared
access network.

\section{Methodology}
\label{sec-3}
The shift from packet-level performance measures to user-perceived ones,
together with user-behavior-based traffic generation, demands a new methodology
for experiments and the analysis of their results. In this section we describe
the details of experimental platform and system models, generation of traffic
and gathering of performance measures, and a framework for a comparative
analysis of the results.

\subsection{Experimental Platform and System Models}
\label{sec-3-1}
Due to the complexity of protocols and the interactive nature of traffic in the
study of network architectures and protocols, researchers now heavily depend on
experiments with simulation and/or test beds implementing proposed architectures
and protocols rather than mathematical analyses under simplifying assumptions.
Especially the experimental platform for this study should be able to capture
the interaction of traffic flows through a complete protocol stack, which are
generated based on user behavior models at the application/session layers. We do
also need a full control of the whole end-to-end network configuration to
eliminate the effect of complicated factors on the performance measures of
interest (e.g., background traffic in metro and backbone networks). For these
reasons, we implemented a virtual test bed composed of detailed simulation
models based on OMNeT++ \cite{Varga:01} and INET framework \cite{INET}, which
provide models for end-user applications as well as a complete TCP/IP protocol
stack.

\begin{figure}[htb]
  \centering
  \includegraphics[angle=-90,width=.8\linewidth]{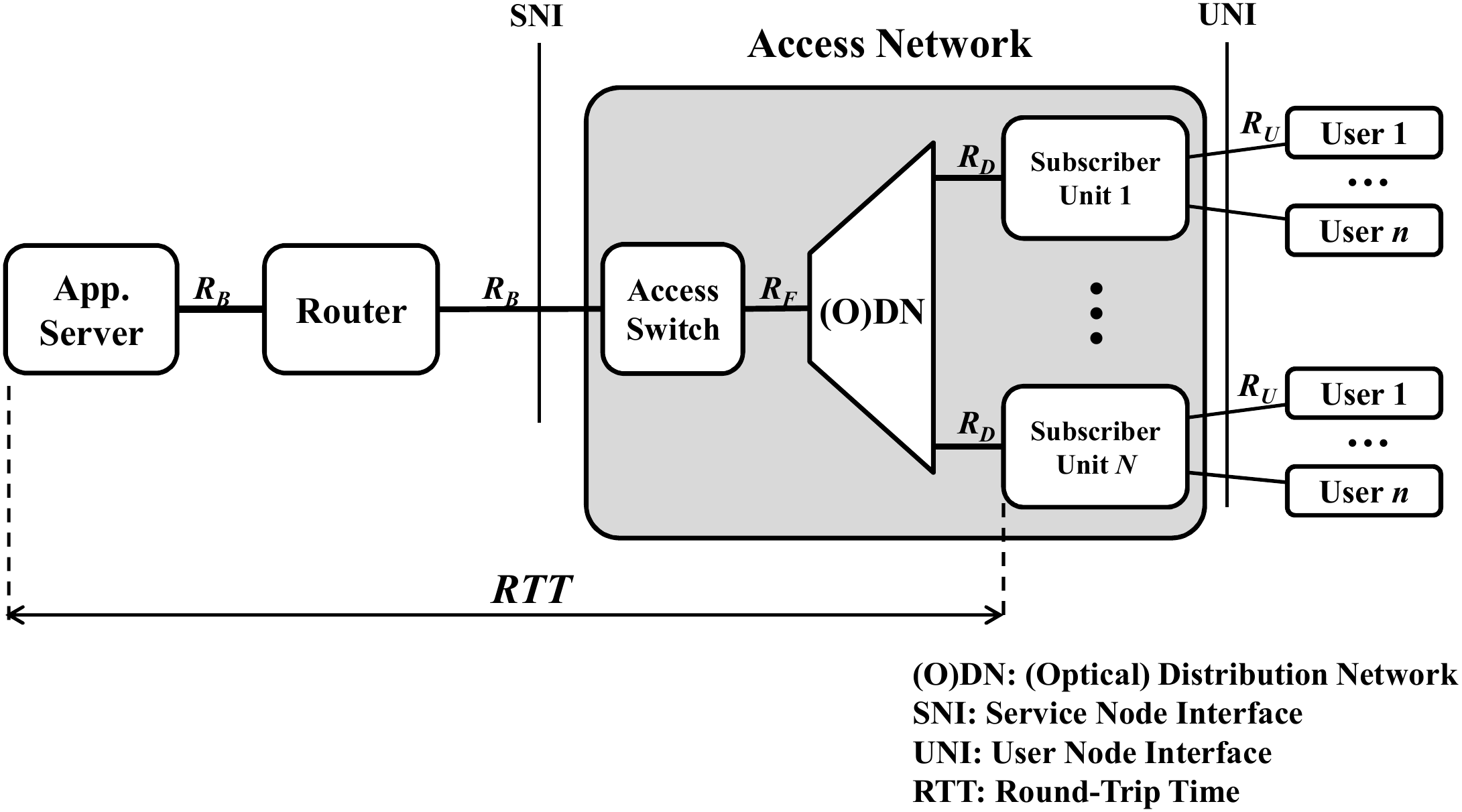}
  \caption{\label{fig:virtual_testbed}An overview of a virtual test bed for a
    shared access network.}
\end{figure}

\begin{figure}[htb]
  \centering
  \includegraphics[angle=-90,width=.8\linewidth]{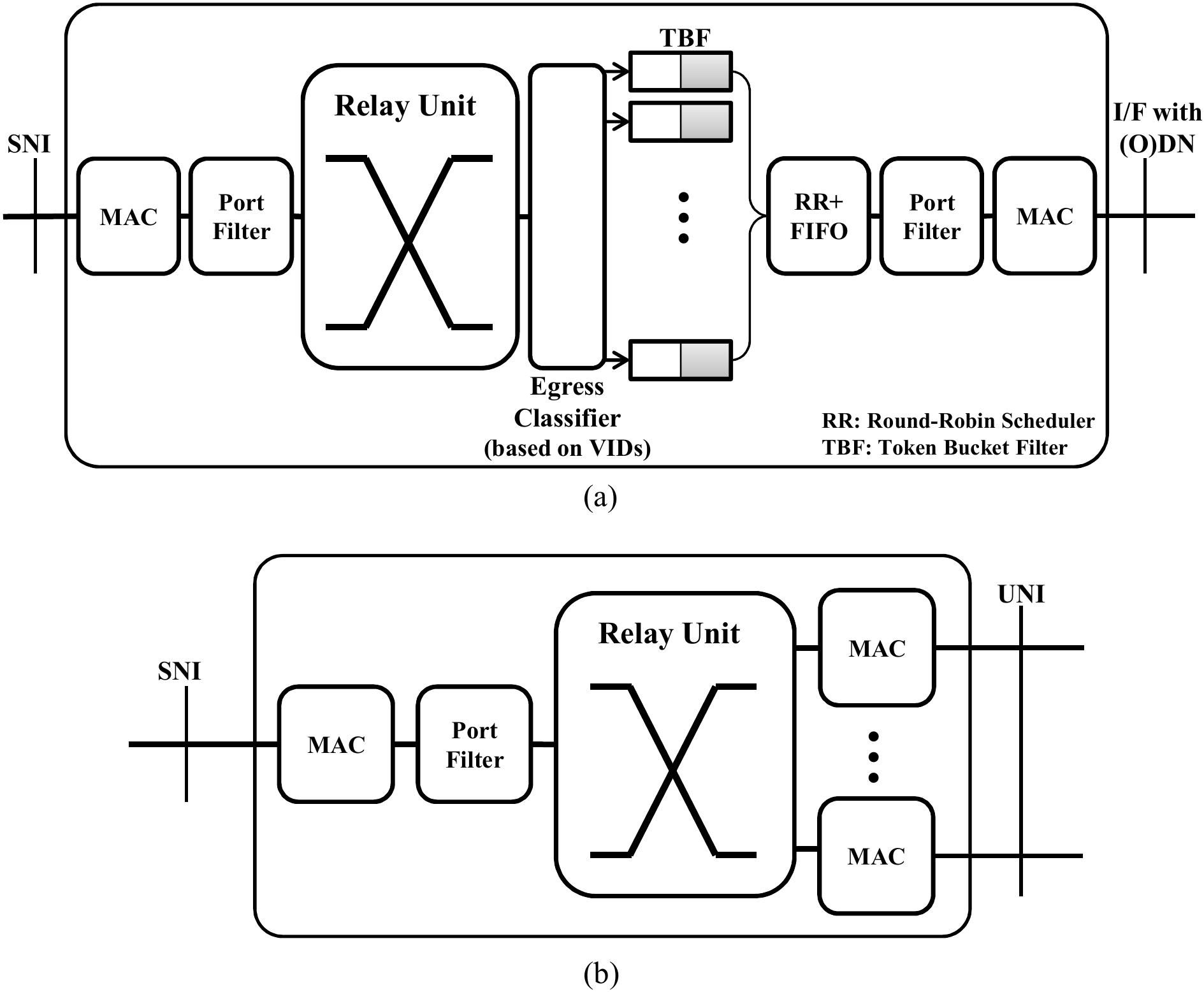}
  \caption{\label{fig:access_components}VLAN-based implementations of (a) the
    access switch and (b) the subscriber unit.}
\end{figure}

\begin{figure}[htb]
  \centering
  \includegraphics[angle=-90,width=.7\linewidth]{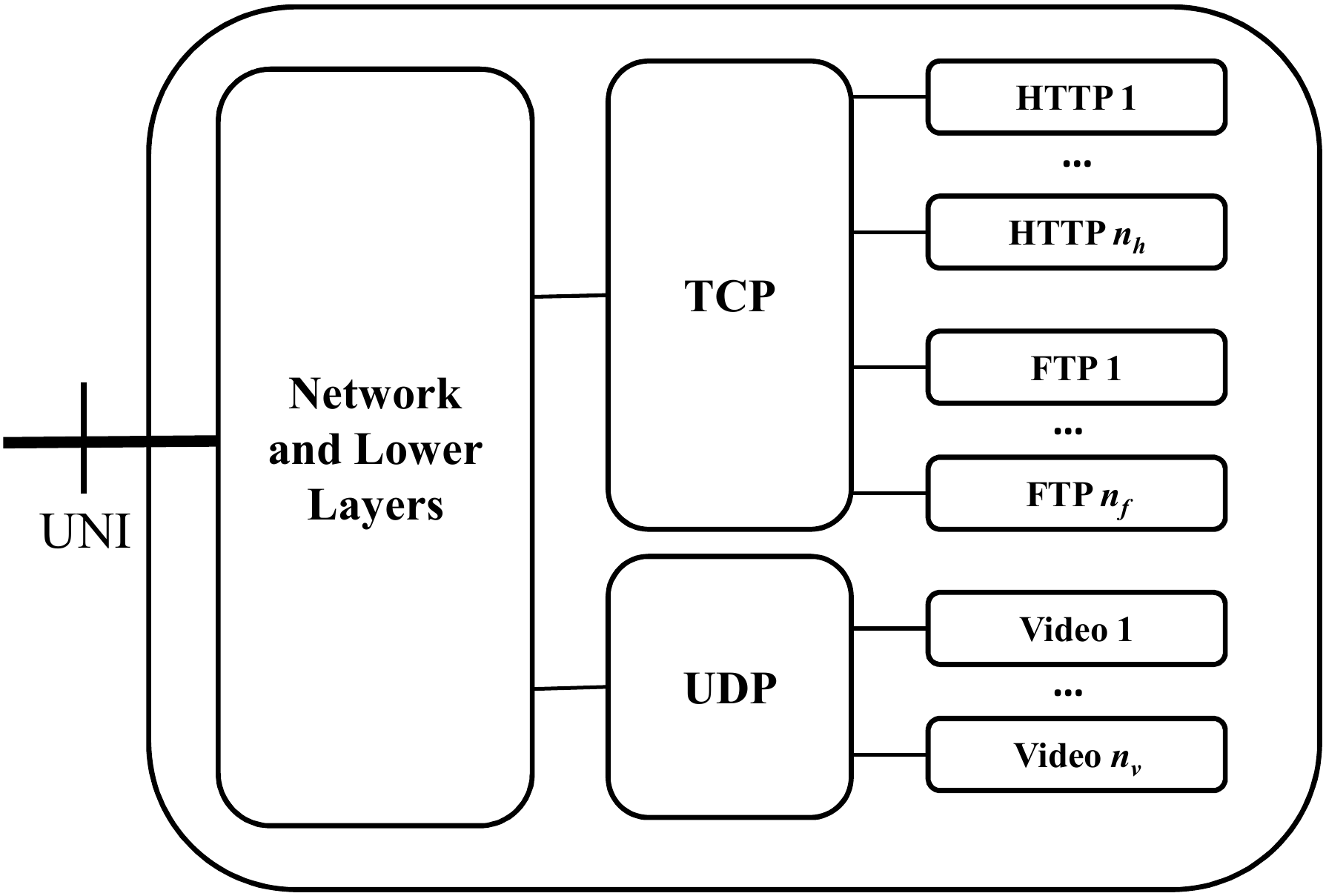}
  \caption{\label{fig:user_model}An end-user model.}
\end{figure}

Fig.\(~\)\ref{fig:virtual_testbed} shows an overview of the virtual test bed for
a shared access network. Virtual local area network (VLAN)-based implementations
of the access switch and the subscriber unit are shown in
Fig.\(~\)\ref{fig:access_components}, which abstract key features essential for
this study from specific systems like the cable modem termination system (CMTS)
and the cable modem for cable Internet and the optical line termination (OLT)
and the optical network unit (ONU) for passive optical networks (PONs). As for
the (optical) distribution network ((O)DN), we use a VLAN-aware Ethernet switch
to model it.  Fig.\(~\)\ref{fig:user_model} also shows a model for an end-user
which is connected to the subscriber unit through the UNI. Note that there could
be multiple \emph{users} who share the broadband connection of a
\emph{subscriber} as shown in
Fig.\(~\)\ref{fig:virtual_testbed}.\footnote{Consider, for example, a household
  (i.e., a subscriber) where family members (i.e., users) share an Internet
  connection through a home network.}

The reason we adopt VLAN-based abstract models for the access switch, the shared
distribution network (i.e., ODN), and the subscriber units is that we need
models which can provide features common to specific systems (e.g., cable
Internet and Ethernet PON (EPON)), while being practical enough to be compatible
with other components and systems of the virtual test bed like the backbone
router and the application server implementing standard protocols (e.g., hyper
text transfer protocol (HTTP) and file transfer protocol (FTP) over TCP/IP). To
identify each subscriber in our shared access model, we assign a unique VLAN
identifier (VID) to each subscriber, which is similar to the service identifier
(SID) in cable Internet and the logical link identifier (LLID) in EPON. The
egress classification in the access switch is based on VIDs and the classified
downstream flows go through TBFs and are scheduled by a round-robin scheduler as
shown in Fig.\(~\)\ref{fig:access_components}. Note that, because we mainly
focus on the performance of downstream traffic in this paper, we do not
consider upstream traffic shaping at the subscriber unit.

\subsection{Traffic Generation and Performance Measurement}
\label{sec-3-2}
\begin{figure*}[htb]
  \centering
  \includegraphics[angle=-90,width=.85\linewidth]{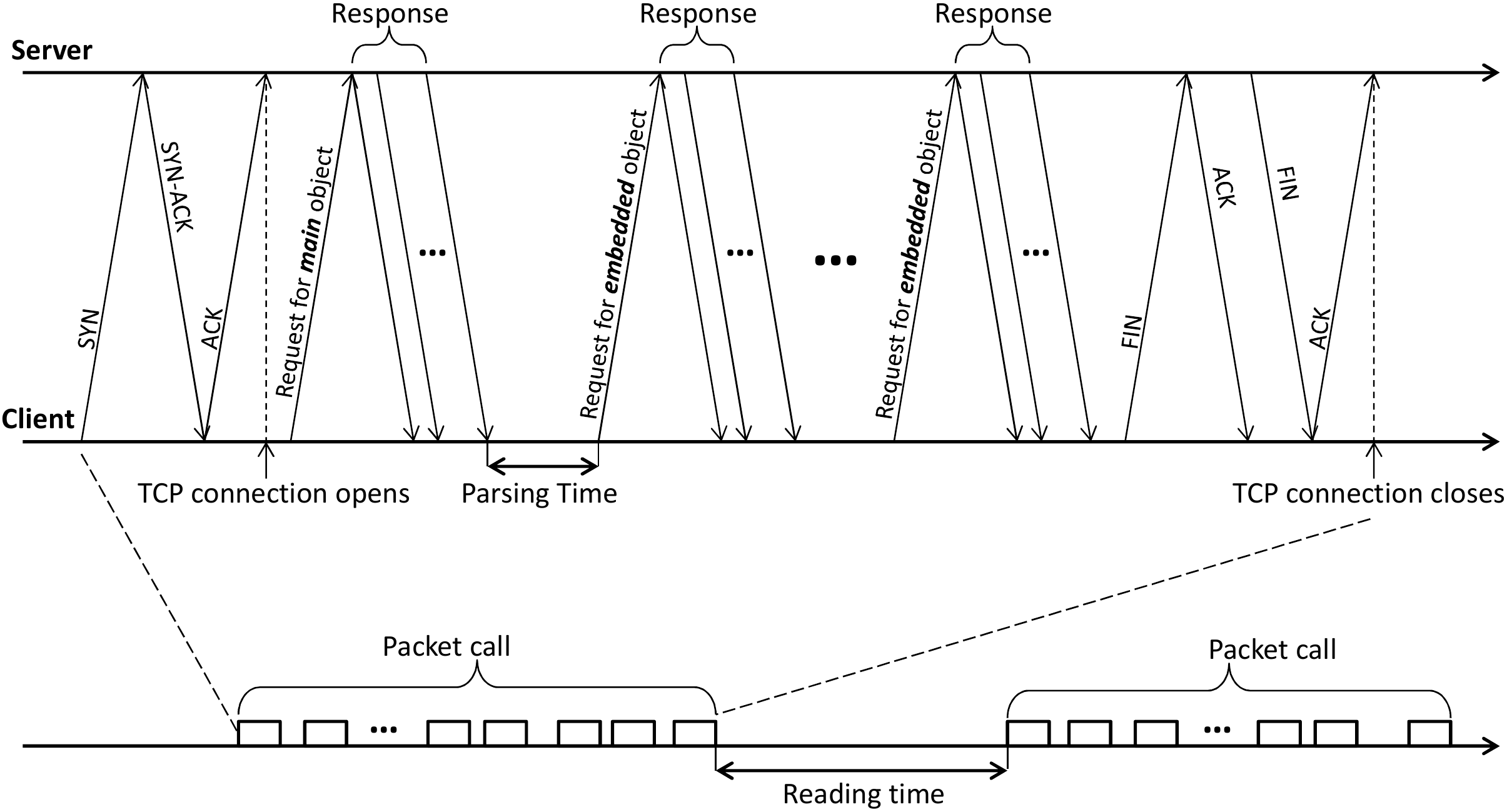}
  \caption{\label{fig:tcp_traffic_model}A user-behavior-based traffic model for
    HTTP and FTP traffic.}
\end{figure*}

As for HTTP and FTP traffic, we use the user-behavior-based traffic model shown
in Fig.\(~\)\ref{fig:tcp_traffic_model}, which is based on the model introduced
by the 3rd generation partnership project (3GPP) for CDMA2000 evaluation
\cite{cdma2000-evaluation:09}.  The FTP traffic is a special case of this model,
where there are a request and response(s) for the main object (e.g., a file to
download) only. The parameter values used for experiments are summarized in
Table \ref{tbl:http_ftp_traffic_parameters}.  Note that these traffic models and
parameter values are gaining wider acceptance among other standard bodies (e.g.,
WiMAX Forum \cite{wimax-sys-eval:08} and IEEE 802.20 \cite{mbwa-evaluation:03})
and now serving as reasonable consensus models bringing some uniformity in
comparisons of systems.

As metrics of user-perceived performance for HTTP and FTP traffic, we collect
packet-call-level performance measures during an experiment. For instance, the
web page delay suggested as the main performance metric for web browsing in
\cite{shankaranarayanan:01} corresponds to the packet call delay defined as the
time taken from the beginning to the end of a packet call in
Fig.\(~\)\ref{fig:tcp_traffic_model}. Likewise, the average page throughput and
the mean page transfer rate in \cite{shankaranarayanan:01} are defined as the
ratio of the mean packet call size (i.e., the size of all objects in a packet
call) to the mean packet call delay and the mean of all the packet call size to
packet call delay ratios, respectively. Later in the comparative analysis, even
though we obtain all three packet call-level metrics for both traffic types, we
mainly use the average packet call delay (i.e., the web page delay) and the
average packet call throughput as main metrics for HTTP and FTP traffic,
respectively.

\begin{table}
  \begin{threeparttable}
    \centering
    \caption{Parameter values for HTTP and FTP traffic models
      \cite{cdma2000-evaluation:09}}
    \label{tbl:http_ftp_traffic_parameters}
    \begin{tabular}{|l|l|} \hline Parameters/Measurements & Best Fit
      (Parameters) \\ \hline\hline
      %%% HTTP %%%
      \multicolumn{2}{|c|}{HTTP Model} \\ \hline
      HTML Object Size [Byte]: & Truncated Lognormal: \\
      Mean=10710, SD=25032, & $\mu$=8.35, $\sigma$=1.37, \\
      Min=100, Max=2M & Min=100, Max=2M \\ \hline
      Embedded Object Size [Byte]: & Truncated Lognormal: \\
      Mean=7758, SD=126168, & $\mu$=6.17, $\sigma$=2.36, \\
      Min=50, Max=2M & Min=50, Max=2M \\ \hline
      Number of Embedded Objects: & Truncated Pareto\tnote{1}~: \\
      Mean=5.64, Max=53 & $\alpha$=1.1, $k$=2, $m$=55 \\ \hline
      Parsing Time [sec]: & Exponential: \\
      Mean=0.13 & $\lambda$=7.69 \\ \hline
      Reading Time [sec]: & Exponential: \\
      Mean=30 & $\lambda$=0.033 \\ \hline
      Request Size [Byte]: & Uniform: \\
      Mean=318.59, SD=179.46 & $a$=0, $b$=700 \\ \hline\hline
      %%% FTP %%%
      \multicolumn{2}{|c|}{FTP Model} \\ \hline
      File Size [Byte]: & Truncated Lognormal: \\
      Mean=2M, SD=0.722M, & $\mu$=14.45, $\sigma$=0.35, \\
      Max=5M & Max=5M \\ \hline
      Reading Time [sec]: & Exponential: \\
      Mean=180 & $\lambda$=0.006 \\ \hline
      Request Size [Byte]: & Uniform: \\
      Mean=318.59, SD=179.46 & $a$=0, $b$=700 \\ \hline
    \end{tabular}
    \begin{tablenotes}
    \item[1] $k$ is subtracted from the generated random value to obtain a
      distribution for the number of embedded objects.
    \end{tablenotes}
  \end{threeparttable}
\end{table}

As for streaming video traffic, we use two video traces for VBR-coded
H.264/advanced video coding (AVC) clips from Arizona state university (ASU)
video trace library \cite{Auwera:08-1} --- i.e., the common intermediate format
(CIF) ``Star Wars IV'' and the high definition (HD) format ``Terminator2'' ---
whose properties are summarized in Table
\ref{tbl:video_traffic_overview}. Frames are encapsulated by real-time transport
protocol (RTP) and then user datagram protocol (UDP), and finally carried as the
payload of IP packets.  The starting frame is selected randomly from the trace
at the beginning of simulation and the whole trace is cycled throughout the
period of the stream. Once all the frames of a given trace have been processed,
the same video is immediately started up, but with a new randomly-selected
starting frame. In this way the resulting traffic is kept random without any
fixed phase relationship among multiple video streams from the same trace file
during the simulation \cite{Seeling:04}.
\begin{table*}[!tbp]
  \begin{threeparttable}
    \centering
    \caption{Properties/statistics of streaming video traffic models
      \cite{Auwera:08-1}}
    \label{tbl:video_traffic_overview}
    \begin{tabular}{|l|l|l|} \hline \multirow{2}{*}{Property/Statistic} &
      \multicolumn{2}{|c|}{Value} \\ \cline{2-3} & lower service rate\tnote{1}~
      & higher service rate\tnote{2}~ \\ \hline
      Video Clip & ``Star Wars IV'' & ``Terminator2'' \\
      Encoding (VBR-coded) & H.264/AVC & H.264/AVC \\
      Encoder & H.264 Full & H.264 FRExt  \\
      Duration & $\sim$30 min & $\sim$10 min \\
      Frame Size & CIF 352x288 & HD 1280x720p \\
      GoP Size & 16 & 12 \\
      Number of B Frames & 3 & 2 \\
      Quantizer & 10 & 10 \\
      Mean Frame Bit Rate & 1.63 Mbit/s & 28.6 Mbit/s \\ \hline
    \end{tabular}
    \begin{tablenotes}
    \item[1] $\leq$ 20 Mbit/s.
    \item[2] $\geq$ 30 Mbit/s.
    \end{tablenotes}
  \end{threeparttable}
\end{table*}

As a metric of user-perceived quality of video stream, we adopt the decodable
frame rate (DFR) which is defined as the ratio of successfully decoded frames at
a receiver to the total number of frames sent by a video source
\cite{Ziviani:05}: The larger the value of DFR, the better the video quality
perceived by the end-user. Note that in calculating DFR, we also take into
account the effect of a 5-second de-jitter buffer as suggested in
\cite{cdma2000-evaluation:09}: Because we know a cumulative display time of each
frame with respect to the first startup I frame thanks to the decoding frame
number in the video trace, we can convert frame delay into frame loss as shown
in Algorithm~\ref{alg:video_delay_loss_conversion}.\footnote{We use this simple
  model of decoding and play-out buffering; detailed modeling like adaptive
  media playout scheme in \cite{kalman04:_adapt} is beyond the scope of this
  paper.}
\begin{algorithm}[hbtp]
  \SetKw{Initialization}{Initialization} \Initialization{}\; $i \longleftarrow$
  decoding frame number of the startup I frame\; $t_i \longleftarrow$ arrival
  time of the startup I frame\; $T_F \longleftarrow$ frame period\tcc*{e.g.,
    33.3 ms} $T_D \longleftarrow$ startup delay\tcc*{e.g., 5 s} \BlankLine
  \SetKw{Arrival}{Arrival} \Arrival{on the arrival of a video frame with a
    decoding frame number $j$ and arrival time $t_{j}$}\; \If{$t_{j} > t_{i} +
    T_{F} \times (j - i) + T_{D}$}{ Discard the arrived frame\; }
  \caption{Frame delay-loss conversion in the video streaming model.}
  \label{alg:video_delay_loss_conversion}
\end{algorithm}

We can also consider an integrated traffic model where both FTP and streaming
video traffic are embedded within HTTP traffic flows once \emph{behavioral
  traffic models} for these cases (especially for embedded streaming video like
YouTube) are available.

For details of the implemented traffic models, readers are referred to
\cite{Kim:11-1}.\footnote{Note that the simulation models, configurations, and
  scripts for pre- and post-processing are available online at
  ``http://github.com/kyeongsoo/inet-hnrl''. }

\subsection{Comparative Analysis Framework}
\label{sec-3-3}
Because our investigation depends on simulation experiments rather than
mathematical analyses as discussed in Sec. \ref{sec-3-1}, we need a way to
systematically take into account the statistical variability in measured data
from the experiments. For a comparative analysis of the effect of ISP traffic
shaping with respect to the unshaped case, we do also need to collectively
process the multiple user-perceived performance metrics for different service
types (i.e., HTTP, FTP, and streaming video) of possibly multiple users
belonging to the same subscriber; note that, while the metrics discussed in
Sec. \ref{sec-3-2} are to capture the performance of individual traffic flows
perceived by a user, the ISP traffic shaping is done at the subscriber level
over the combined traffic flows from the multiple users within the subscriber.

\begin{figure}[htb]
  \centering
  \includegraphics[angle=-90,width=.8\linewidth]{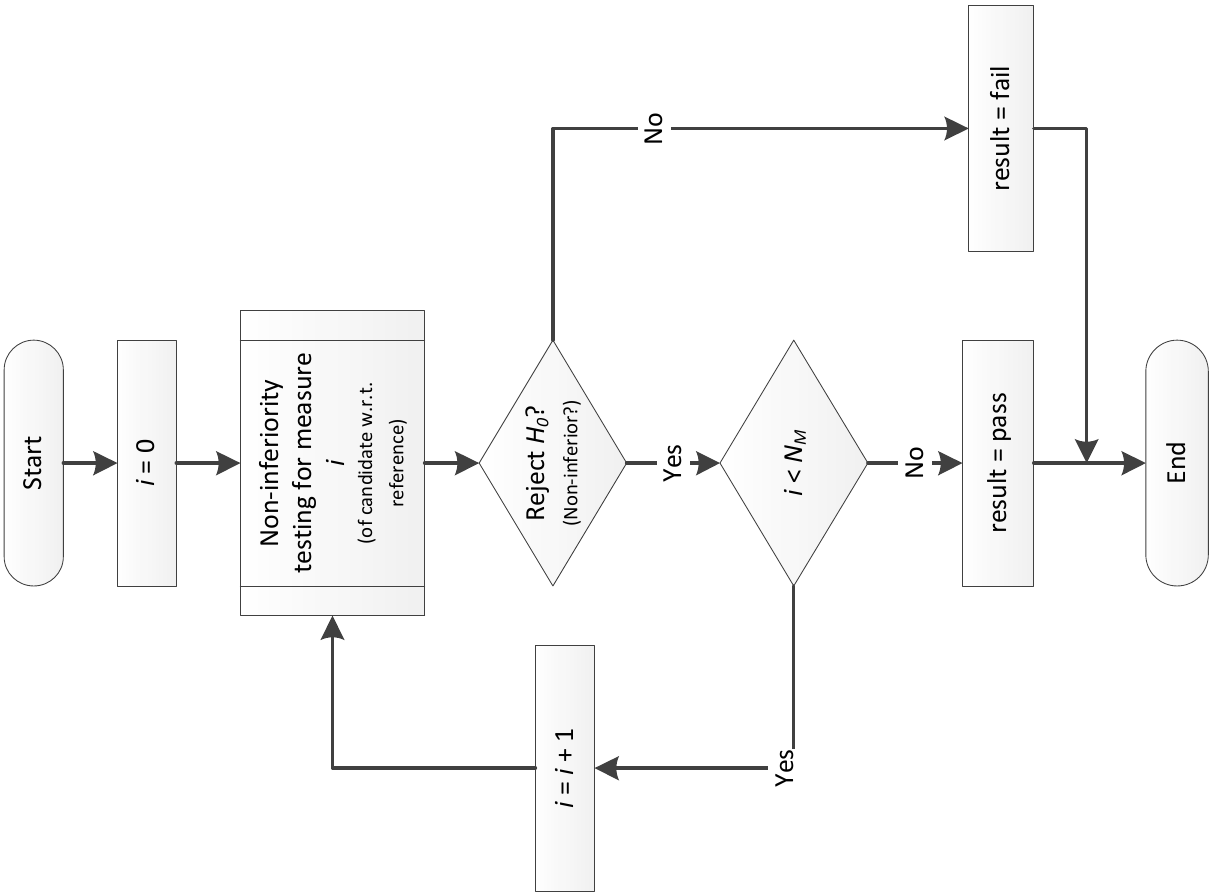}
  \caption{An overview of multivariate non-inferiority testing procedure
    \cite{Kim:11-4}.}
  \label{fig:multivariate_noninferiority_testing}
\end{figure}

To take into account the statistical variability in measured data and process
multiple performance metrics in an integrated way during the comparison, we
adopt the comparative analysis framework that we proposed for the clean-slate
design of next-generation optical access in \cite{Kim:11-4}. In this framework
we use the user-perceived performance of a single subscriber under no traffic
shaping as a reference case against which we compare the user-perceived
performance of all other shaped configurations either with a single subscriber
or with multiple subscribers. Then, using the multivariate non-inferiority
testing procedure shown in Fig.~\ref{fig:multivariate_noninferiority_testing},
we find system configurations (e.g., the number of subscribers and the number of
traffic flows per subscriber) and TBF parameter values (e.g., token bucket size
and token generation rate) for which the user-perceived performance are
statistically \emph{non-inferior} to those of the reference case.

The comparative analysis based on user-perceived performance with respect to a
reference case is inspired by the equivalent circuit rate (ECR) measure for a
quantitative comparison of hybrid fiber coaxial (HFC) cable-based shared access
network and DSL-based dedicated access network architectures in
\cite{shankaranarayanan:01}. The original comparison framework for the ECR
assumes that each \emph{active} user is using a single application, i.e., web
browsing, and bases the comparison on the user-perceived performance of that
application only, where there is no differentiation between the user and the
subscriber and no statistical comparison procedure is used to take into account
the statistical variability in measured data.

The comparative analysis framework proposed in \cite{Kim:11-4} addresses these
drawbacks of the original ECR framework by extending it to multiple performance
metrics and taking into account statistical variability of measured data in
comparison using multivariate non-inferiority testing. The non-inferiority
testing is a one-sided variant of the equivalence testing used in Medicine and
Biology for the establishment of the equivalence between two different clinical
trials or drugs \cite{Berger:96}. The non-inferiority testing procedure is based
on statistical hypothesis testing and as such takes into account the statistical
variability in measured data. To compare multiple performance metrics in an
integrated way, the non-inferiority testing is extended by intersection-union
testing (IUT) \cite{Berger:96}. In this way we can formally define the
equivalence of the results from two configurations (i.e., user-perceived
performance in our case).

\begingroup
\setlength{\medmuskip}{0mu} \setlength{\thinmuskip}{0mu}
\setlength{\thickmuskip}{0mu} In the multivariate non-inferiority testing shown
in Fig.\(~\)\ref{fig:multivariate_noninferiority_testing}, the null and the
alternative hypotheses of the non-inferiority testing for each measure $M_{i}$
(e.g., web page delay), $i=1,~2,\ldots,~N_{M}$, are given by
\endgroup
\begin{equation}
  \left\{
    \begin{array}{l}
      H_{0}:~\mu_{i,C} - \mu_{i,R} \geq \delta_{i} \\
      H_{1}:~\mu_{i,C} - \mu_{i,R} < \delta_{i}
    \end{array}
  \right.
  \label{eq:hypotheses}
\end{equation}
where $\mu_{i,C}$ and $\mu_{i,R}$ denote population means of $M_{i}$ for the
candidate (i.e., shaped) and the reference (i.e., unshaped) configurations,
respectively, and $\delta_{i}$ represents the tolerance for the measure
$M_{i}$. The null hypothesis ($H_0$) is rejected if the limit of one-sided
confidence interval for the difference (i.e., $\mu_{i,C} - \mu_{i,R}$) is less
than the tolerance \cite{silva09:_method_equiv_nonin_testin}. This means that
the result from the candidate configuration is ``at least as good as'' the
reference one for the given measure $M_{i}$. Note that for each measure $M_{i}$,
we need to determine an appropriate tolerance value ($\delta_{i}$) and, if
needed, change the direction of inequalities accordingly. For example, we need
to change the hypotheses for the packet call throughput of FTP traffic (unlike
delay, higher throughput is better) as follows:
\begin{equation}
  \left\{
    \begin{array}{l}
      H_{0}:~\mu_{i,C} - \mu_{i,R} \leq -\delta_{i} \\
      H_{1}:~\mu_{i,C} - \mu_{i,R} > -\delta_{i}
    \end{array}
  \right.
  \label{eq:alternative_hypotheses}
\end{equation}

For details of the comparative analysis framework, readers are referred to
\cite{Kim:11-4}.

\section{Simulation Results}
\label{sec-4}
%%% to begin the group for math mode spacing adjustment
\begingroup
\setlength{\medmuskip}{0mu} \setlength{\thinmuskip}{0mu}
\setlength{\thickmuskip}{0mu}
%%%
The experiment configurations considered in this paper --- i.e., two unshaped
(i.e., U$_{1}$ and U$_{2}$) and eighteen shaped (i.e., S$_{1,1}$--S$_{1,9}$ and
S$_{2,1}$--S$_{2,9}$) ones --- are summarized in Table
\ref{tbl:experiment_config}: For all the configurations, the backbone line rate
($R_{B}$) and the end-to-end round-trip time ($RTT$) are fixed to 100 Gbit/s and
10 ms, respectively. As for the access line rate for the distribution ($R_{D}$)
and the feeder ($R_{F}$), two values of 100 Mbit/s and and 1 Gbit/s are
considered, which could represent the capacities provided by the cable Internet
(i.e., DOCSIS 3.0) and the EPON, respectively. Unless stated otherwise, the
number of HTTP streams ($n_{h}$), the number of FTP streams ($n_{f}$), and the
number of video streams ($n_{v}$) per user are set to 1.
\begin{table*}[!tbp]
  \begin{threeparttable}
    \centering
    \caption{Summary of experiment configurations}
    \label{tbl:experiment_config}
    \begin{tabular}{|c|r|r|r|r|r|} \hline \multirow{2}{*}{Config.} &
      \multicolumn{3}{|c|}{Network Parameters} & \multicolumn{2}{|c|}{TBF
        Parameters} \\ \cline{2-6} & \multicolumn{1}{|c|}{$RTT$} &
      \multicolumn{1}{|c|}{$R_{B}$} & \multicolumn{1}{|c|}{$R_{F}$, $R_{D}$,
        $R_{U}$} & \multicolumn{1}{|c|}{TGR\tnote{1}~} &
      \multicolumn{1}{|c|}{TBS\tnote{2}~} \\ \hline\hline U$_{1}$ &
      \multirow{20}{*}{10 ms} & \multirow{20}{*}{100 Gbit/s} &
      \multirow{10}{*}{100 Mbit/s} & \multicolumn{2}{|c|}{Unshaped} \\
      \cline{1-1}\cline{5-6} S$_{1,1}$ & & & & \multirow{3}{*}{2 Mbit/s} & 1 MB\tnote{3} \\
      \cline{1-1}\cline{6-6} S$_{1,2}$ & & & & & 10 MB \\ \cline{1-1}\cline{6-6}
      S$_{1,3}$ & & & & & 100 MB \\ \cline{1-1}\cline{5-6} S$_{1,4}$ & & & &
      \multirow{3}{*}{10 Mbit/s} & 1 MB \\ \cline{1-1}\cline{6-6} S$_{1,5}$ & & &
      & & 10 MB \\ \cline{1-1}\cline{6-6} S$_{1,6}$ & & & & & 100 MB \\
      \cline{1-1}\cline{5-6} S$_{1,7}$ & & & & \multirow{3}{*}{20 Mbit/s} & 1 MB
      \\ \cline{1-1}\cline{6-6} S$_{1,8}$ & & & & & 10 MB \\
      \cline{1-1}\cline{6-6} S$_{1,9}$ & & & & & 100 MB \\
      \cline{1-1}\cline{4-4}\cline{5-6} U$_{2}$ & & & \multirow{10}{*}{1 Gbit/s} &
      \multicolumn{2}{|c|}{Unshaped} \\ \cline{1-1}\cline{5-6} S$_{2,1}$ & & & &
      \multirow{3}{*}{30 Mbit/s} & 10 MB \\ \cline{1-1}\cline{6-6} S$_{2,2}$ & & &
      & & 100 MB \\ \cline{1-1}\cline{6-6} S$_{2,3}$ & & & & & 1 GB\tnote{4} \\
      \cline{1-1}\cline{5-6} S$_{2,4}$ & & & & \multirow{3}{*}{60 Mbit/s} & 10 MB
      \\ \cline{1-1}\cline{6-6} S$_{2,5}$ & & & & & 100 MB \\
      \cline{1-1}\cline{6-6} S$_{2,6}$ & & & & & 1 GB \\ \cline{1-1}\cline{5-6}
      S$_{2,7}$ & & & & \multirow{3}{*}{90 Mbit/s} & 10 MB \\
      \cline{1-1}\cline{6-6} S$_{2,8}$ & & & & & 100 MB \\
      \cline{1-1}\cline{6-6} S$_{2,9}$ & & & & & 1 GB \\ \hline
    \end{tabular}
    \begin{tablenotes}
    \item [1] Token generation rate.
    \item [2] Token bucket size.
    \item [3] 1 MB = 10$^{6}$ bytes.
    \item [4] 1 GB = 10$^{9}$ bytes.
    \end{tablenotes}
  \end{threeparttable}
\end{table*}

As for the token generation rate (i.e., the long-term average of the service
rate per subscriber), we consider the values of 2 Mbit/s, 10 Mbit/s, and 20
Mbit/s for the access line rate of 100 Mbit/s and the values of 30 Mbit/s, 60
Mbit/s, and 90 Mbit/s for the access line rate of 1 Gbit/s.\footnote{For the
  values of token generation rates, we referred to the current Virgin Media
  Cable traffic management policy \cite{virgin_media_traffic_management}. }; the
minimum 2-Mbit/s service rate is chosen especially because it is a target rate
for the \emph{Universal Service Broadband Commitment} in the Digital Britain
Final Report \cite{department09:_digit_britain_final_repor}. Note that the
combined traffic generation rate of the three flows for HTTP, FTP, and streaming
video traffic from a single user, which is measured at the physical layer
without traffic shaping during preliminary simulations, is 1.83 Mbit/s for
configurations U$_{1}$ and S$_{1,1}$--S$_{1,9}$ with the ``Star Wars IV'' clip
and 30 Mbit/s for configurations U$_{2}$ and S$_{2,1}$--S$_{2,9}$ with the
``Terminator2'' clip. As for the token bucket size, we consider three values of
1 MB, 10 MB, and 100 MB for the access line rate of 100 Mbit/s and ten times
those values for the access line rate of 1 Gbit/s. The peak rate of TBF is set
to the access line rate except for the cases of investigating its effect
discussed in Sec. \ref{sec-4-1}.

During the comparative analysis between the unshaped and the shaped
configurations, we fix the values of the network parameters and the number of
users per subscriber (i.e., $n$) for both configurations, while we vary the
values of TBF parameters and the number of subscribers (i.e., $N$) for the
shaped configurations to investigate the effect of ISP traffic shaping at the
subscriber and the access levels.

Each simulation is run for 3 hours with a warmup period of 20 minutes, both in
simulation time. To calculate confidence intervals and obtain test statistics
for the multivariate non-inferiority testing, each simulation run is repeated
ten times with different random number seeds.

\subsection{With a Single Subscriber}
\label{sec-4-1}
We first investigate the effect of ISP traffic shaping on the user-perceived
performance at the subscriber level with a single subscriber. The amount of
incoming traffic to the TBF is controlled by the number of users per subscriber
$n$. During the investigation the major focus is put on token bucket sizes
which, for a given token generation rate, can provide user-perceived
performance non-inferior to those of a subscriber under no traffic shaping.

Figs.\(~\)\ref{fig:ss_user_performance_100M} and
\ref{fig:ss_user_performance_1G} show representative metrics of user-perceived
performance for a single subscriber, where we observe that the effect of token
bucket size is prominent for both HTTP and the FTP traffic; as for the DFR of
streaming video, the effect of token bucket size is negligible for all the
configurations with the access line rate of 100 Mbit/s (except for token bucket
size of 1 MB), while it is rather significant with the access line rate of 1
Gbit/s where the ratio of video traffic to the total traffic is significantly
higher. The large effect of token bucket size on file transfer performance is
what we can expect from the discussions in \cite{bauer11:_power}, but we found
out that the effect of token bucket size on the average HTTP page delay is also
quite significant as the combined traffic generation rate approaches to the
token generation rate (e.g., $n=1$ for token generation rate of 2 Mbit/s, $n=5$
for 10 Mbit/s, and $n=10$ for 20 Mbit/s); even in such a condition, however, we also
note that the large token bucket size --- i.e., 100 MB and 1 GB for access line
rates of 100 Mbit/s and 1 Gbit/s, respectively --- can provide user-perceived
performance comparable to those without traffic shaping.
%%%
%%% For 100 Mbit/s
%%%
% 1st part of the figure
\begin{figure}[!tpb]
  \begin{subfigure}{\textwidth}
    \centering
    \includegraphics*[width=.9\linewidth]{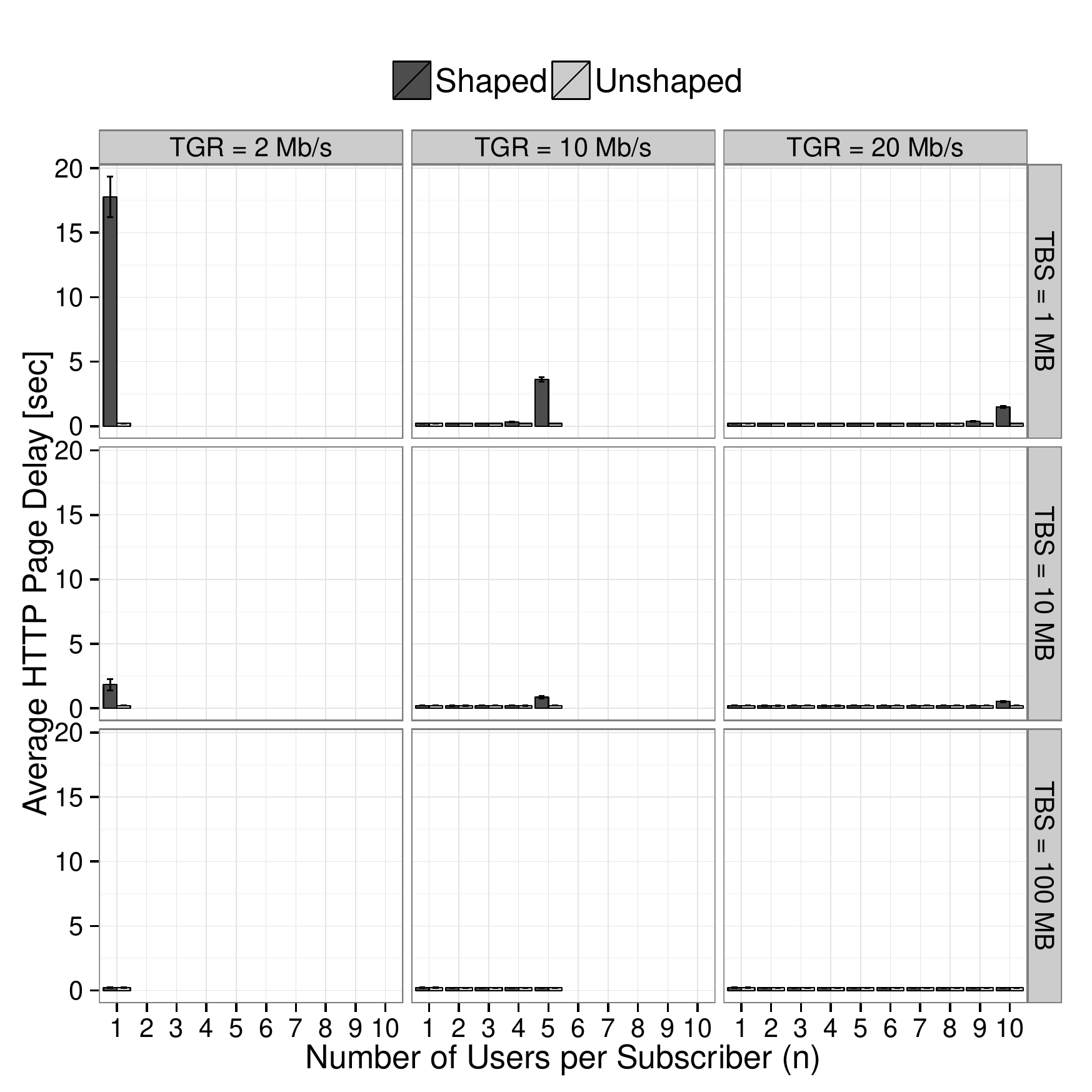}
    \caption{}
  \end{subfigure}
  %   % {\scriptsize (a)}\\
  % % \end{center}
  % % \begin{minipage}{.499\linewidth}
  % %   \begin{center}
  %     \includegraphics*[width=\linewidth]{ss_dr100M-ftp-thr_facet.pdf}\\
  %     {\scriptsize (b)}\\
  % %   \end{center}
  % % \end{minipage}
  % % \hfill
  % % \begin{minipage}{.499\linewidth}
  % %   \begin{center}
  %     \includegraphics*[width=\linewidth]{ss_dr100M-video-dfr_facet.pdf}\\
  %     {\scriptsize (c)}
  %   \end{center}
  % % \end{minipage}
  \caption{User-perceived performance metrics with 95 percent confidence
    intervals for a single subscriber with access line rate of 100 Mbit/s: (a)
    Average session delay of HTTP traffic, (b) average session throughput of
    FTP traffic, and (c) decodable frame rate (Q) of UDP streaming video.}
  \label{fig:ss_user_performance_100M}
\end{figure}
% 2nd part of the figure
\begin{figure}[!tpb]
  \ContinuedFloat
  \begin{subfigure}{\textwidth}
    \centering
    \includegraphics*[width=.9\linewidth]{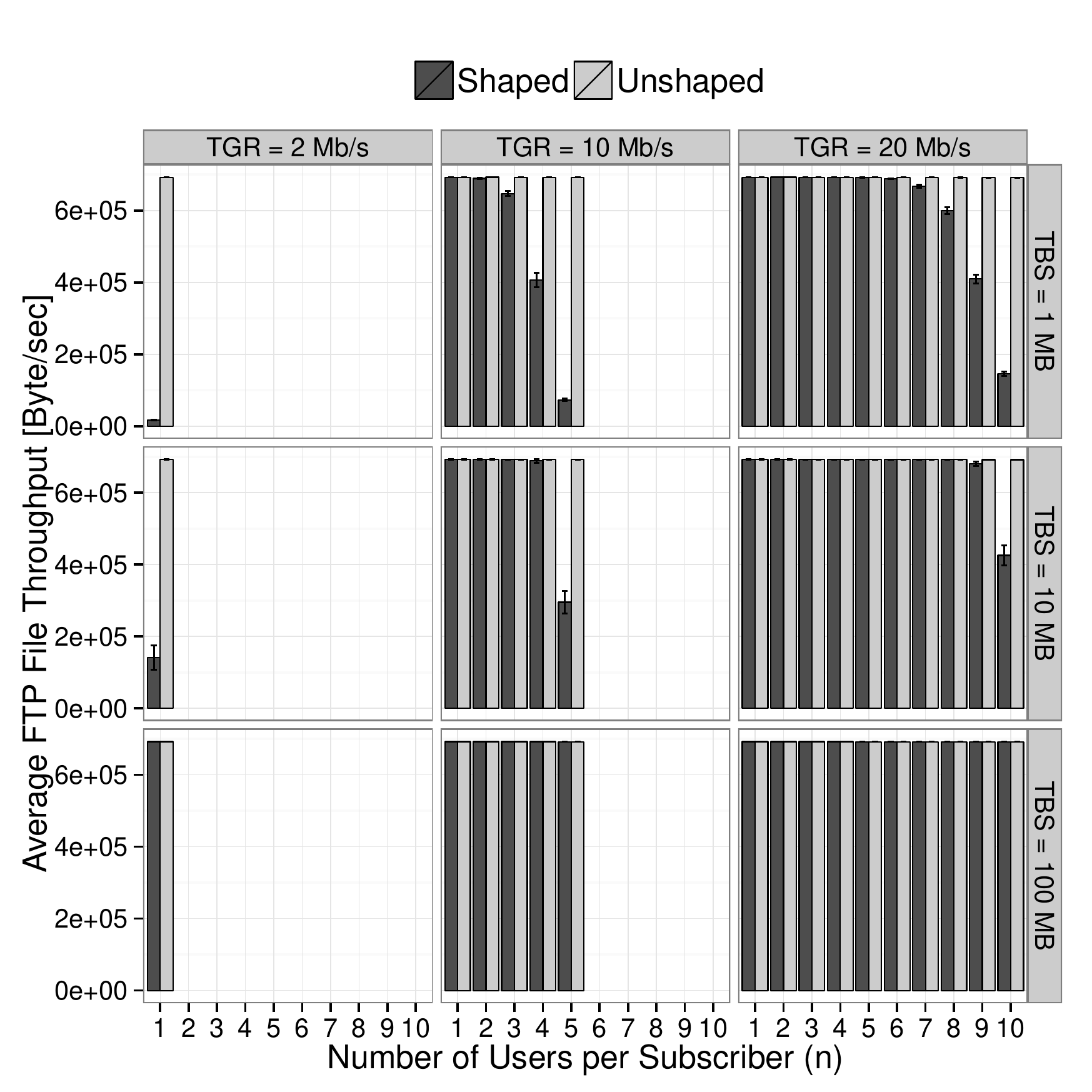}
    \caption{}
  \end{subfigure}
  \caption{User-perceived performance metrics with 95 percent confidence
    intervals for a single subscriber with access line rate of 100 Mbit/s: (a)
    Average session delay of HTTP traffic, (b) average session throughput of
    FTP traffic, and (c) decodable frame rate (Q) of UDP streaming video.}
\end{figure}
% 3rd part of the figure
\begin{figure}[!tpb]
  \ContinuedFloat
  \begin{subfigure}{\textwidth}
    \centering
    \includegraphics*[width=.9\linewidth]{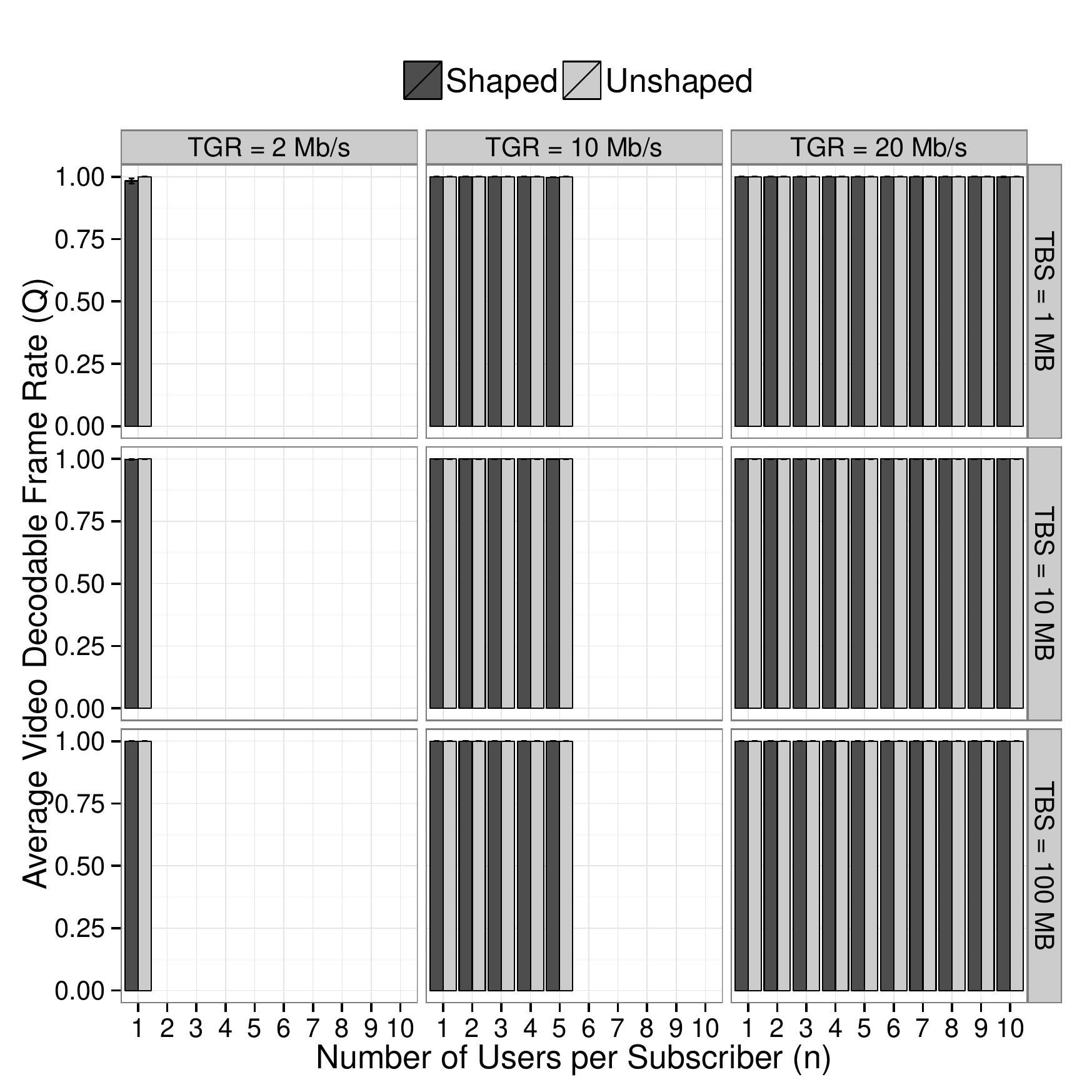}
    \caption{}
  \end{subfigure}
  \caption{User-perceived performance metrics with 95 percent confidence
    intervals for a single subscriber with access line rate of 100 Mbit/s: (a)
    Average session delay of HTTP traffic, (b) average session throughput of
    FTP traffic, and (c) decodable frame rate (Q) of UDP streaming video.}
\end{figure}
%%%
%%% For 100 Mbit/s
%%%
%% 1st part of the figure
\begin{figure*}[!tpb]
  \begin{subfigure}{\textwidth}
    \centering
    \includegraphics*[width=.9\linewidth]{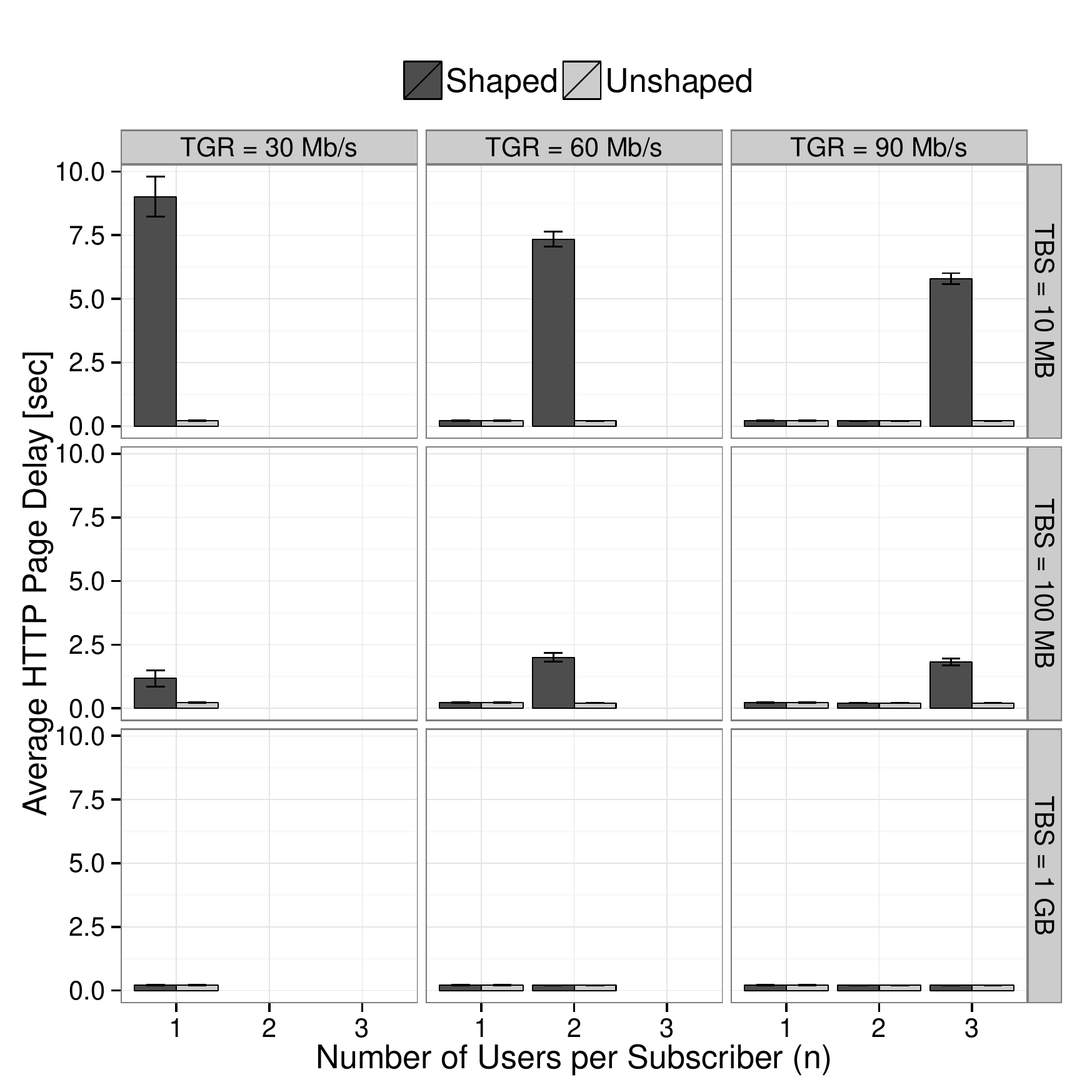}
    \caption{}
  \end{subfigure}
  \caption{User-perceived performance metrics with 95 percent confidence
    intervals for a single subscriber with access line rate of 1 Gbit/s: (a)
    Average session delay of HTTP traffic, (b) average session throughput of FTP
    traffic, and (c) decodable frame rate (Q) of UDP streaming video.}
  \label{fig:ss_user_performance_1G}
\end{figure*}
%% 2nd part of the figure
\begin{figure*}[!tpb]
  \ContinuedFloat
  \begin{subfigure}{\textwidth}
    \centering
    \includegraphics*[width=.9\linewidth]{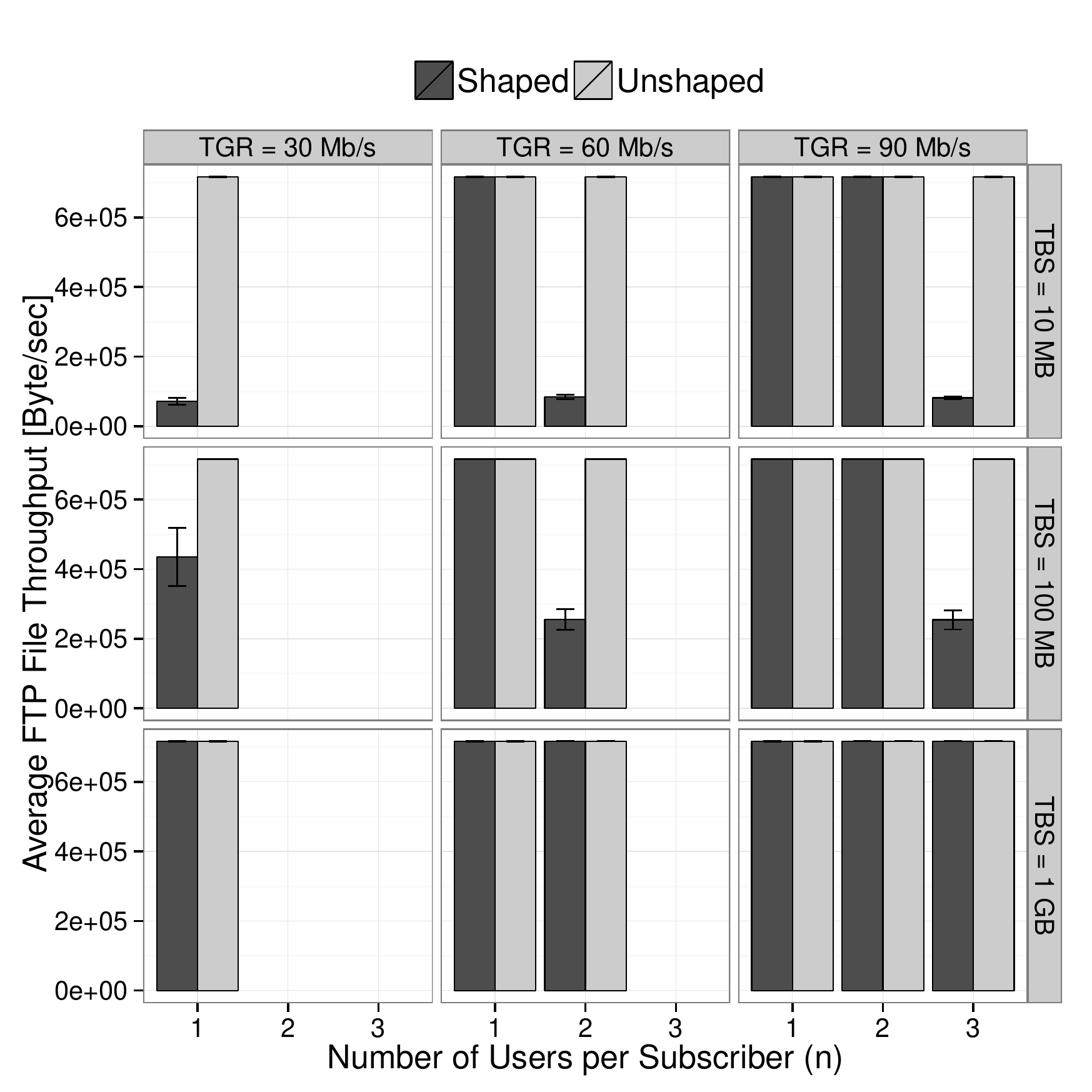}
    \caption{}
  \end{subfigure}
  \caption{User-perceived performance metrics with 95 percent confidence
    intervals for a single subscriber with access line rate of 1 Gbit/s: (a)
    Average session delay of HTTP traffic, (b) average session throughput of FTP
    traffic, and (c) decodable frame rate (Q) of UDP streaming video.}
\end{figure*}
%% 3rd part of the figure
\begin{figure*}[!tpb]
  \ContinuedFloat
  \begin{subfigure}{\textwidth}
    \centering
    \includegraphics*[width=.9\linewidth]{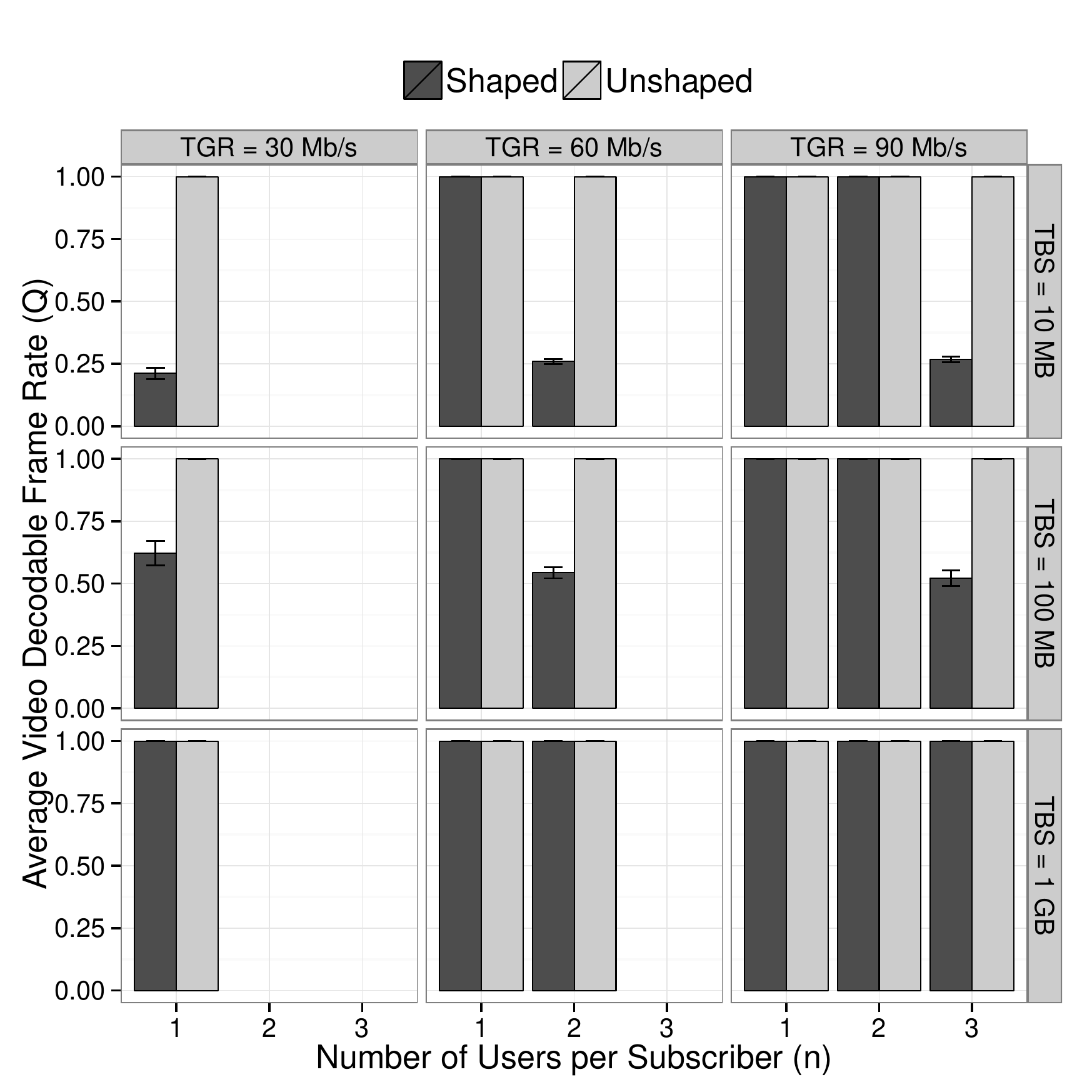}
    \caption{}
  \end{subfigure}
  \caption{User-perceived performance metrics with 95 percent confidence
    intervals for a single subscriber with access line rate of 1 Gbit/s: (a)
    Average session delay of HTTP traffic, (b) average session throughput of FTP
    traffic, and (c) decodable frame rate (Q) of UDP streaming video.}
\end{figure*}

For a comparative analysis, we carried out the multivariate non-inferiority
testing described in Sec. \ref{sec-3-3} for the shaped configurations (i.e.,
S$_{1,1}$--S$_{1,9}$ and S$_{2,1}$--S$_{2,9}$) with respect to the corresponding
reference configurations (i.e., U$_{1}$ and U$_{2}$). We set the tolerance
(i.e., $\delta_i$) to 10 percent of the sample mean of performance measure for
the reference case and the significance level --- i.e., the probability of
rejecting the null hypothesis $H_0$ when it is true
\cite[Section~8.1.2]{Montgomery:94} --- to 0.05. The results are shown in
Fig.\(~\)\ref{fig:ss_user_noninferiority}, where $max(n_{eqv})$ is defined as
the maximum number of users per subscriber (i.e., $n$) of a shaped configuration
which provides user-perceived performance non-inferior to those of the
corresponding unshaped configuration given the same number of users per
subscriber and the access line rate.
\begin{figure*}[!tpb]
  % \begin{minipage}{.499\linewidth}
    \begin{center}
      \includegraphics*[width=.6\linewidth]{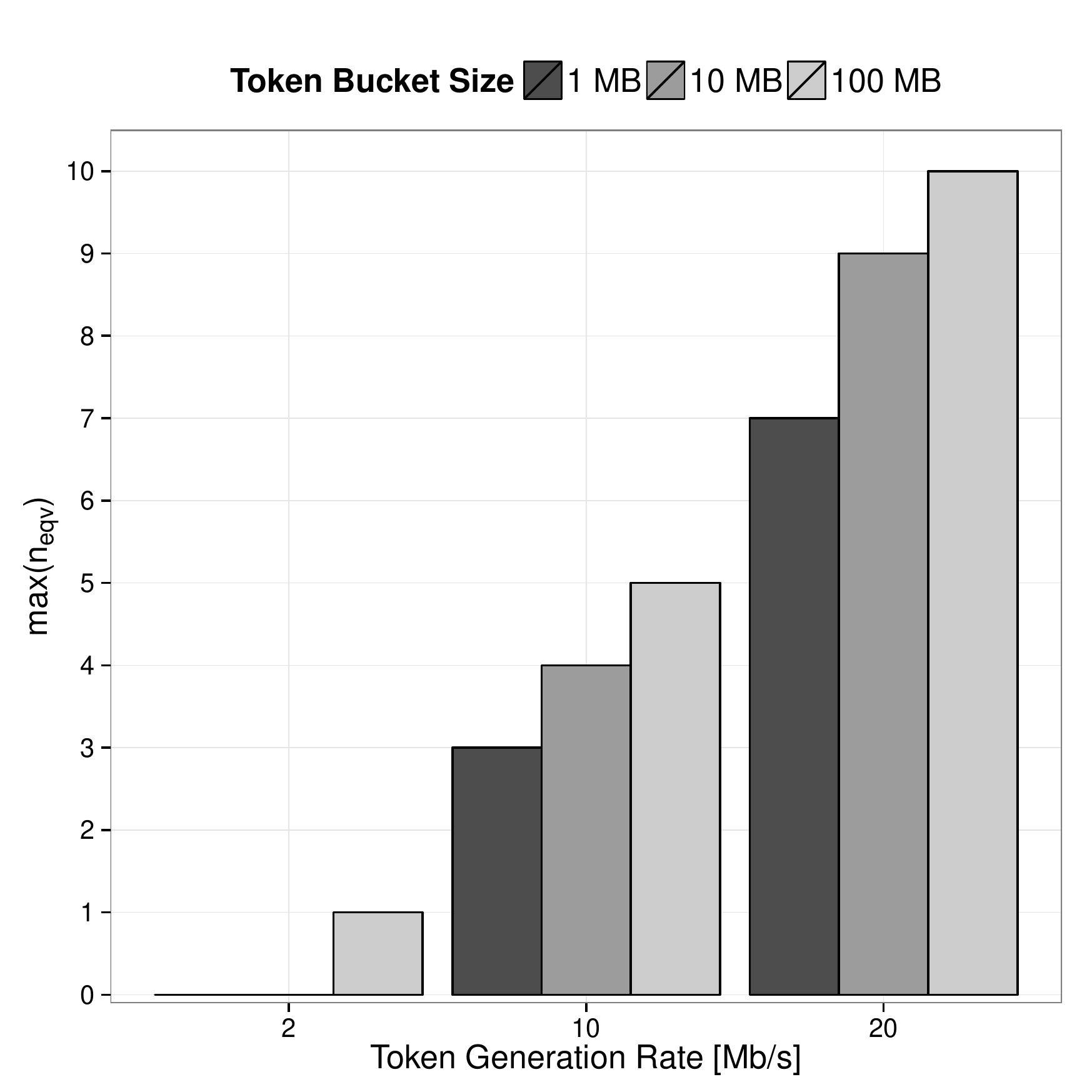}\\
      {\scriptsize (a)}\\
    % \end{center}
  % \end{minipage}
  % \hfill
  % \begin{minipage}{.499\linewidth}
  %   \begin{center}
      \includegraphics*[width=.6\linewidth]{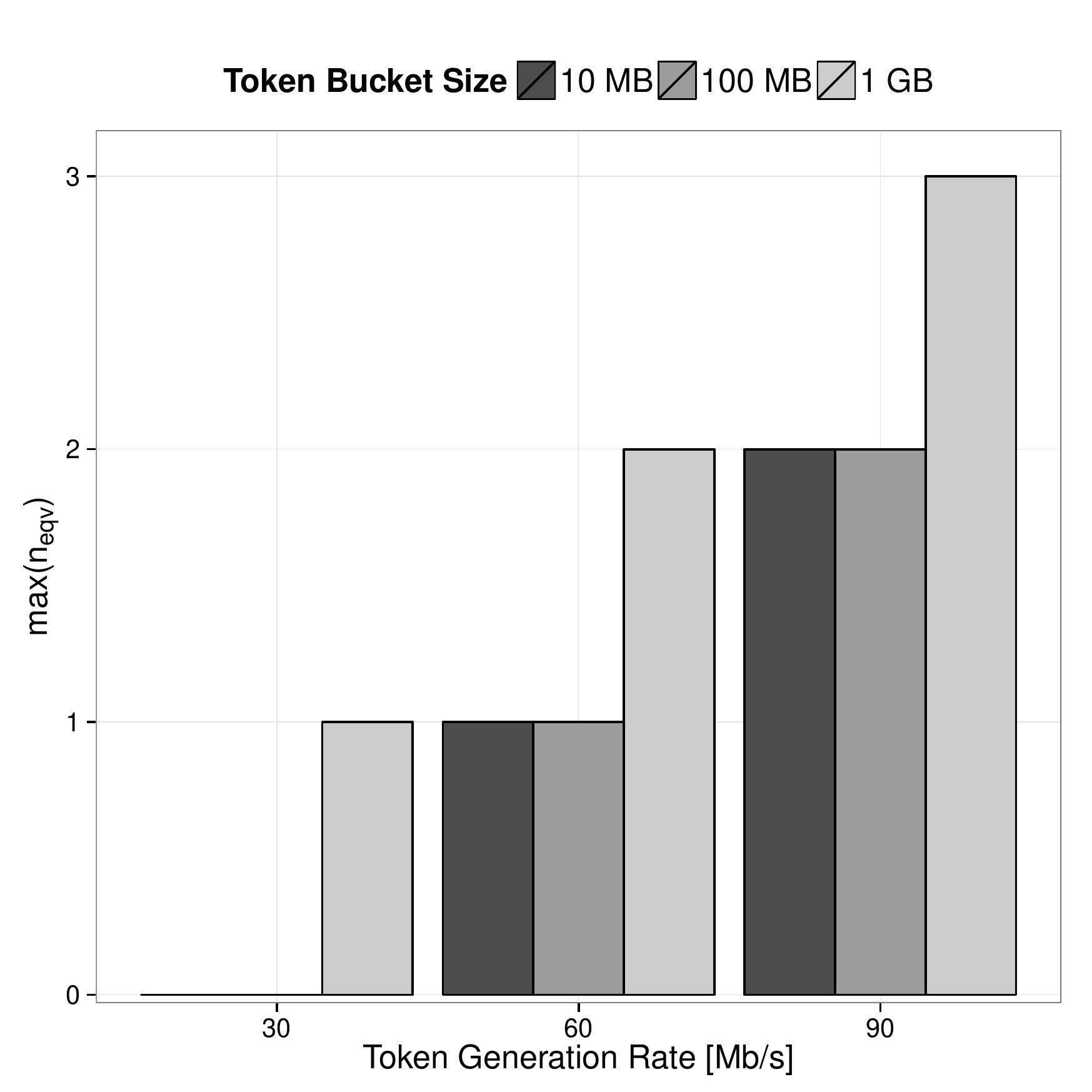}\\
      {\scriptsize (b)}
    \end{center}
  % \end{minipage}
    \caption{Maximum number of users per subscriber of shaped configurations
      % ($max(n_{eqv})$)
      which provides user-perceived performance non-inferior to that of the
      corresponding unshaped configurations for a single subscriber with access
      line rate of (a) 100 Mbit/s and (b) 1 Gbit/s.}
  \label{fig:ss_user_noninferiority}
\end{figure*}

The results for the access line rate of 100 Mbit/s in
Fig.\(~\)\ref{fig:ss_user_noninferiority} (a) show that for the token generation
rate of 2 Mbit/s, the token bucket size of 100 MB can support one user with
user-perceived performance non-inferior to those without traffic shaping, while
for the token generation rates of 10 Mbit/s and 20 Mbit/s, the same token bucket
size can support up to five and ten users respectively. Note that the traffic
from one, five, and ten users per subscriber fully loads the TBF with the token
generation rates of 2 Mbit/s, 10 Mbit/s, and 20 Mbit/s. It is remarkable to see that
with the token bucket size of 100 MB, the token generation rate of mere 2 Mbit/s
can provide user-perceived performance nearly equivalent to those with the
access line rate of 100 Mbit/s, the rate fifty times higher than the token
generation rate. Similar observations are made for the results for the access
line rate of 1 Gbit/s in Fig.\(~\)\ref{fig:ss_user_noninferiority} (b).

With two configurations S$_{1,3}$ (i.e., token generation rate of 2 Mbit/s and
token bucket size of 100 MB for access line rate of 100 Mbit/s) and S$_{2,3}$
(i.e., token generation rate of 30 Mbit/s and token bucket size of 1 GB for
access line rate of 1 Gbit/s) which can support up to one user per subscriber
with user-perceived performance non-inferior to those of unshaped configuration
U$_1$ and U$_2$ respectively, we investigated the effect of the peak rate on
user-perceived performance as shown in
Fig.\(~\)\ref{fig:ss_user_performance_peak_rate}, where we vary the peak rate
from 2 Mbit/s to 100 Mbit/s for S$_{1,3}$ and from 30 Mbit/s to 1 Gbit/s for
S$_{2,3}$. The results show that the peak rate has a significant impact on the
user-perceived performance until it increases to five times the average rate for
both line rates. As for the streaming video, we found out that, when the peak
rate is reduced to the average rate, the resulting decrease in performance is
more significant for S$_{2,3}$ than S$_{1,3}$ because the combined rate of
traffic flows for S$_{2,3}$ is more close to the average rate than that of
S$_{1,3}$. We carried out the investigation of the effect of the peak rate with
other configurations as well and observed similar results.
\begin{figure*}[!tpb]
  \begin{minipage}{.49\linewidth}
    \begin{center}
      \includegraphics*[width=.85\linewidth]{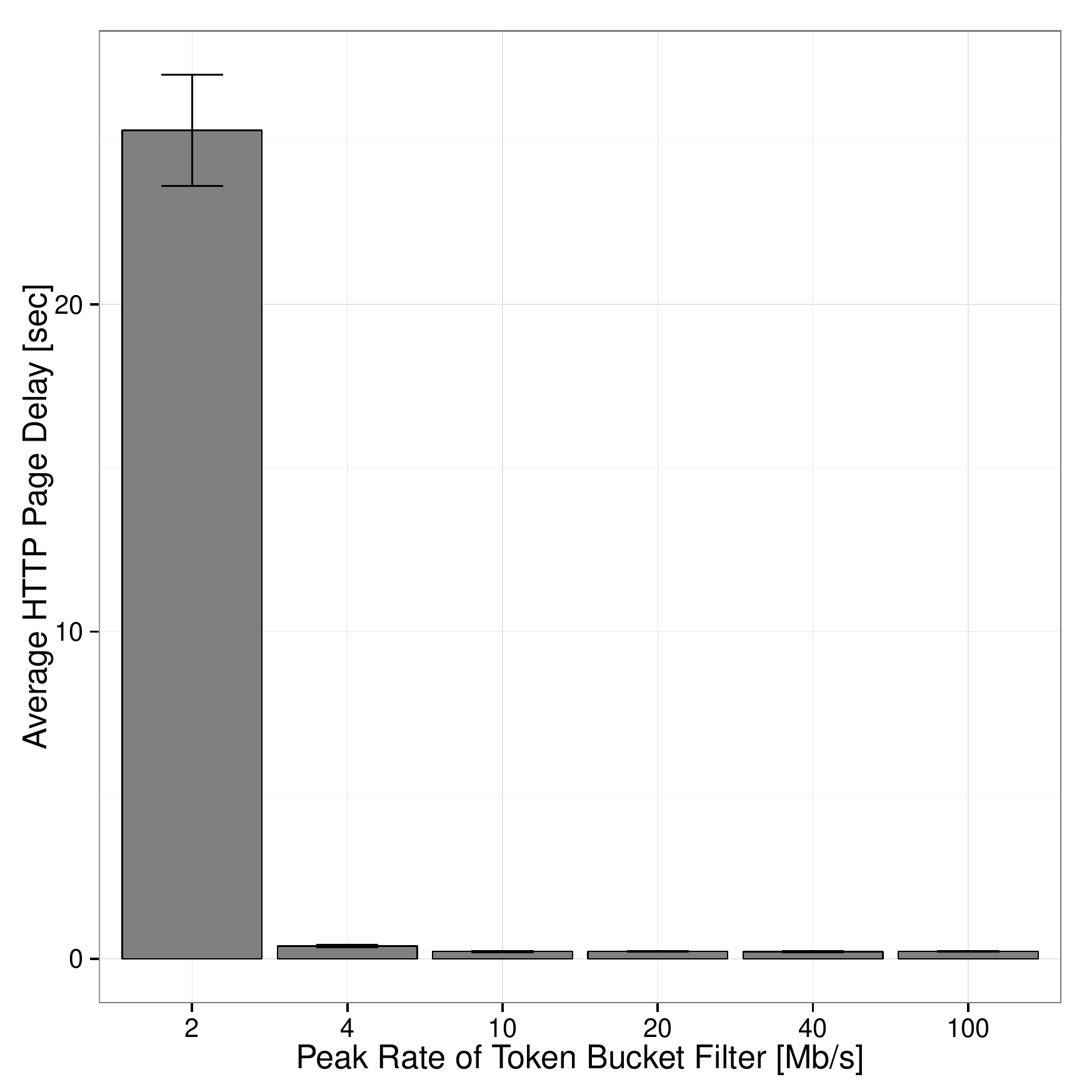}\\
      \includegraphics*[width=.85\linewidth]{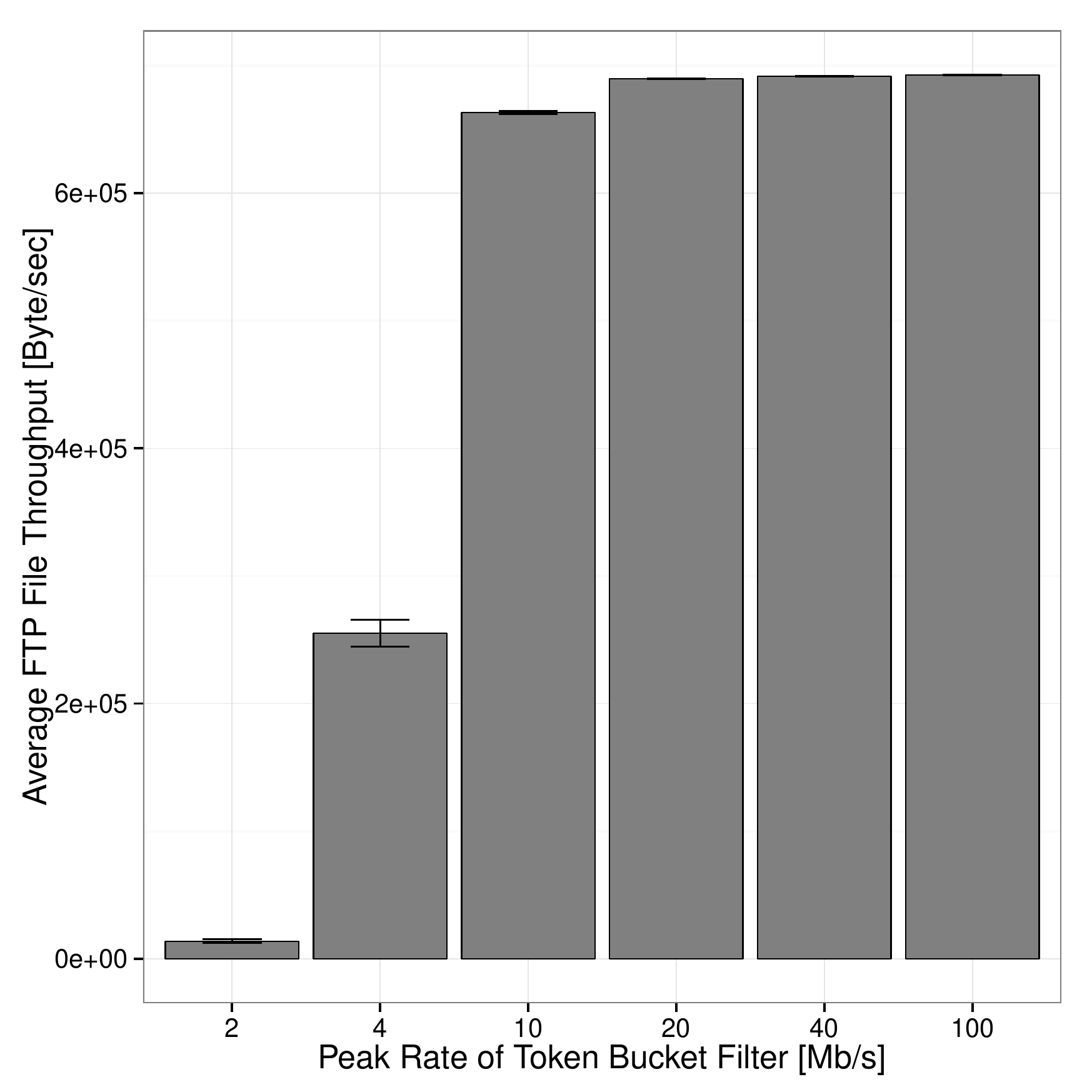}\\
      \includegraphics*[width=.85\linewidth]{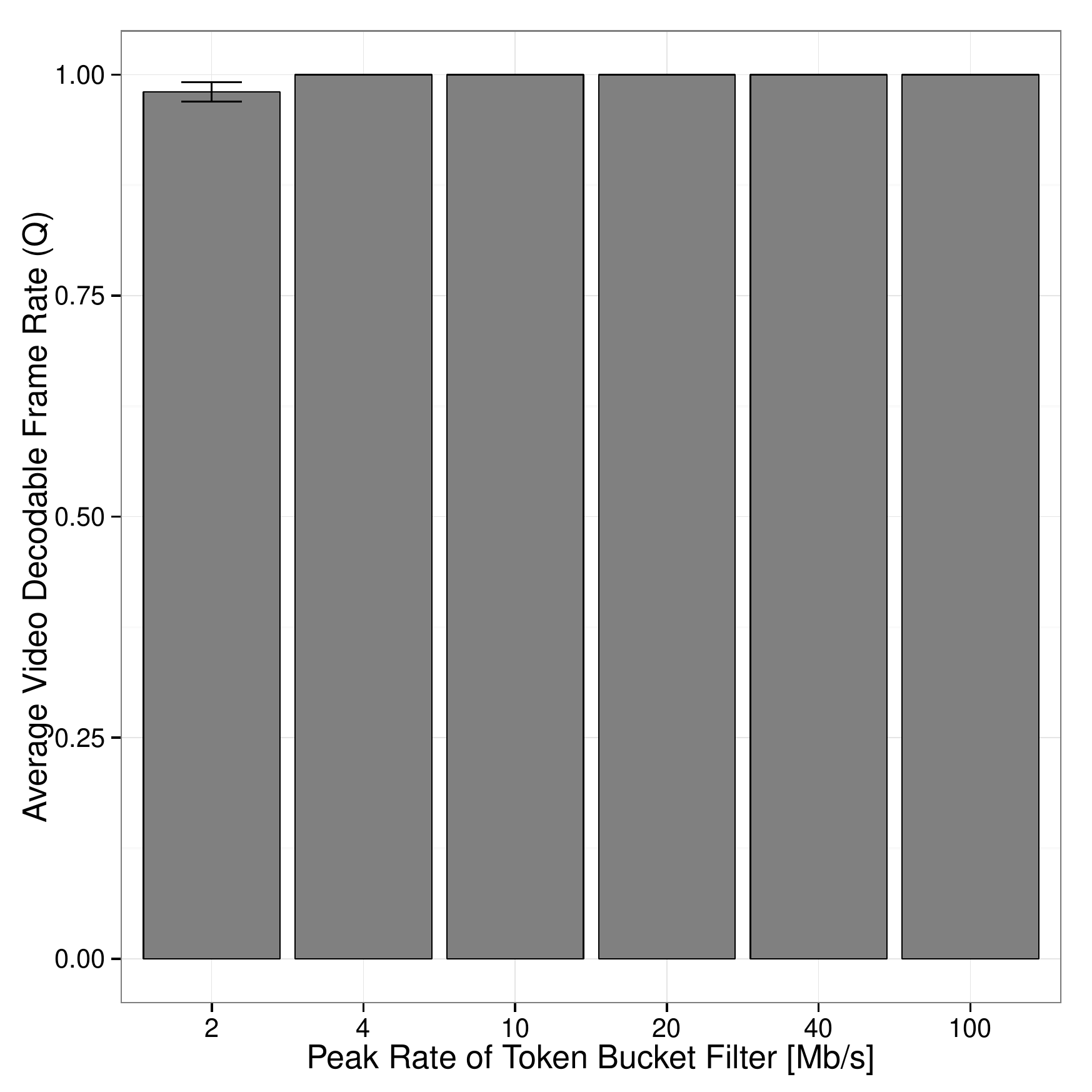}\\
      {\scriptsize (a)}
    \end{center}
  \end{minipage}
  \hfill
  \begin{minipage}{.49\linewidth}
    \begin{center}
      \includegraphics*[width=.85\linewidth]{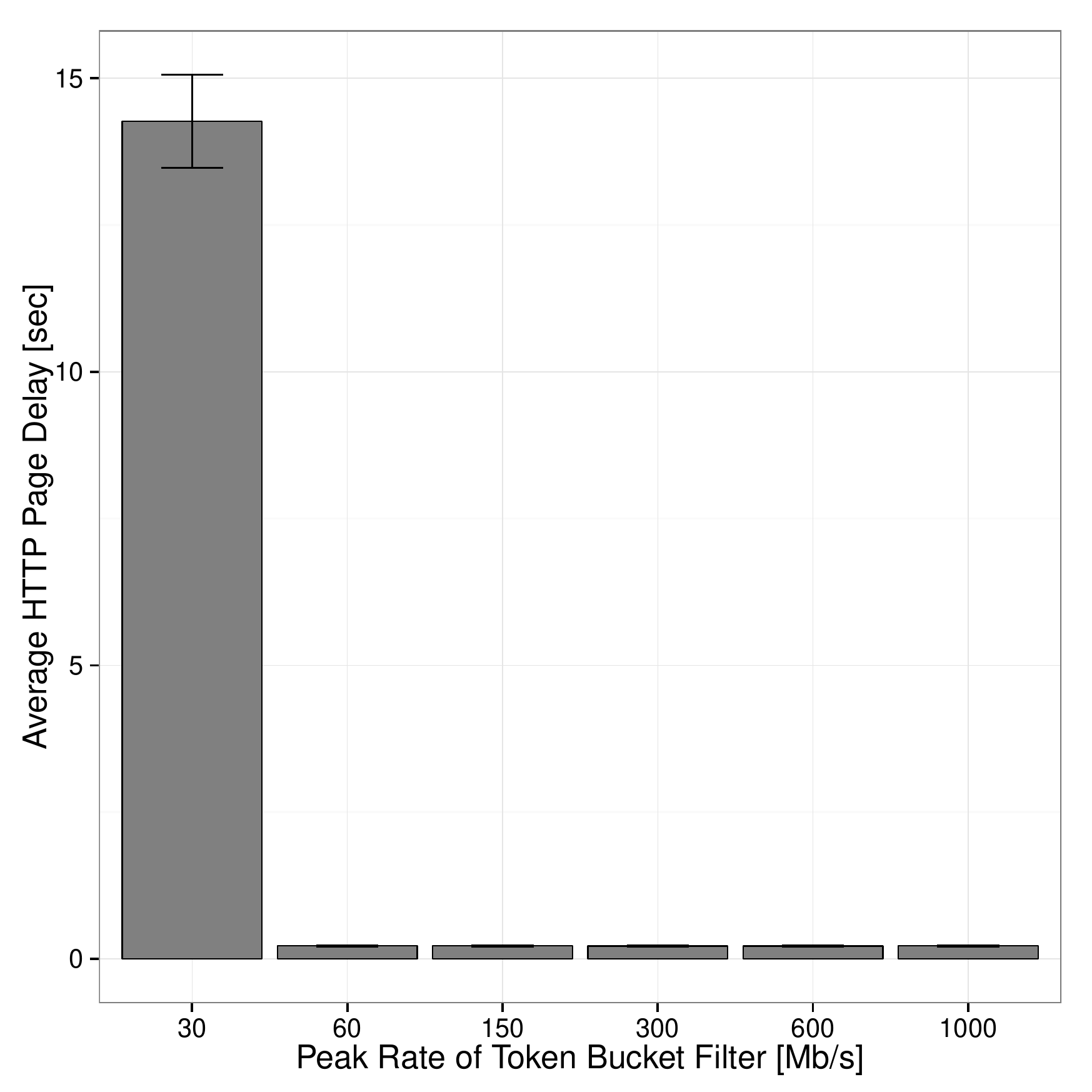}\\
      \includegraphics*[width=.85\linewidth]{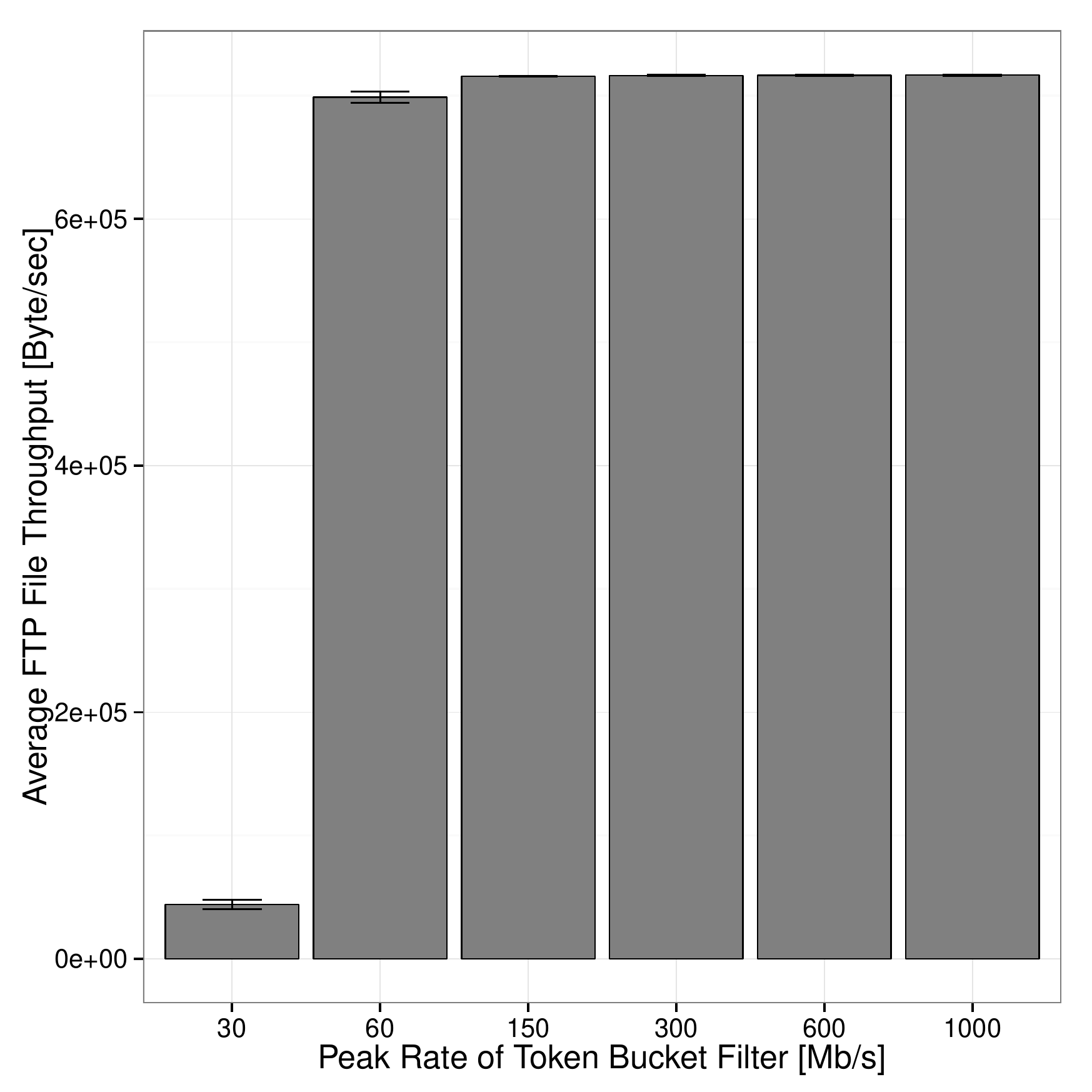}\\
      \includegraphics*[width=.85\linewidth]{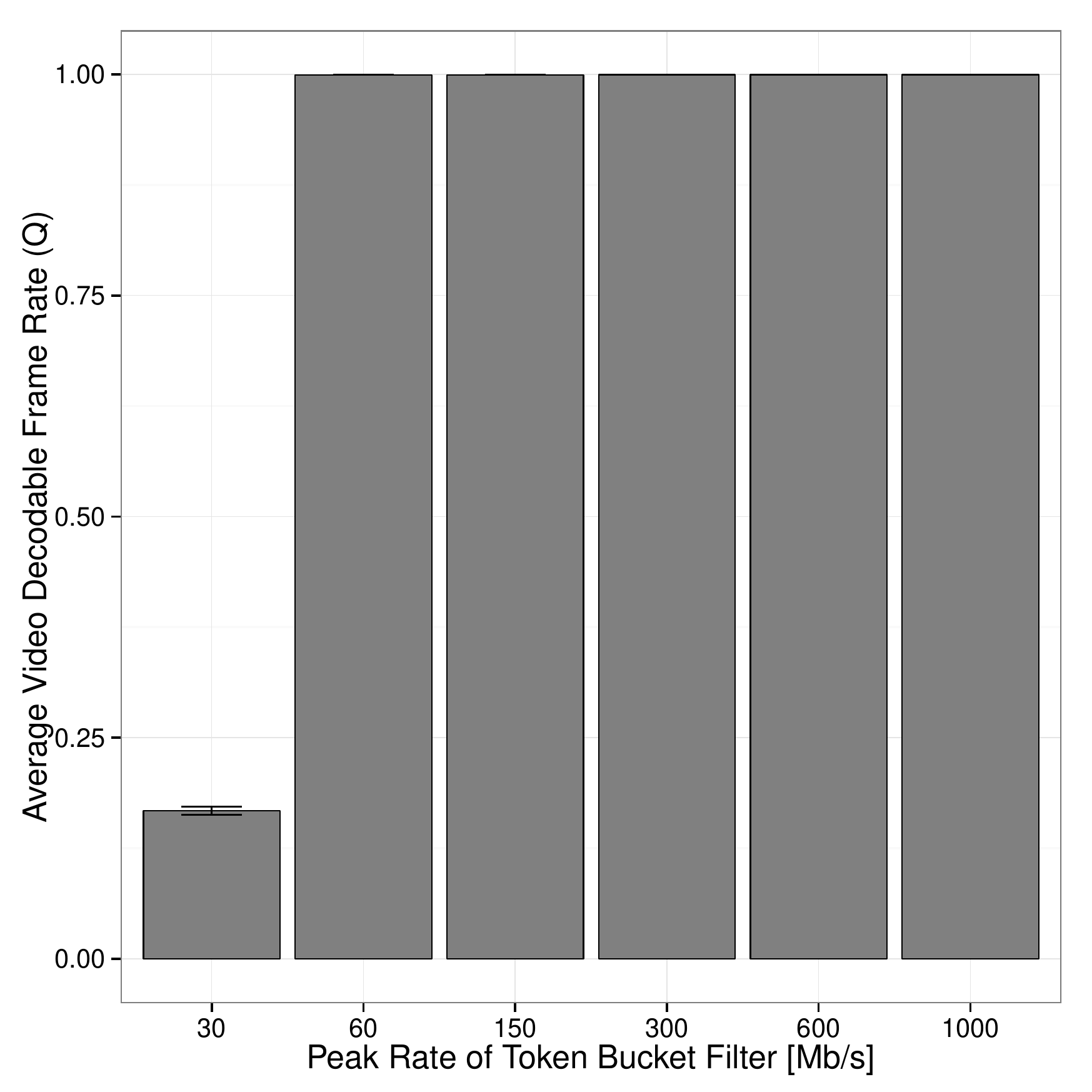}\\
      {\scriptsize (b)}
    \end{center}
  \end{minipage}
  \caption{Effect of the peak rate on user-perceived performance with a single
    subscriber and a singe user: (a) access line rate of 100 Mbit/s with token
    generation rate of 2 Mbit/s and token bucket size of 100 MB and (b) access
    line rate of 1 Gbit/s with token generation rate of 30 Mbit/s and token bucket
    size of 1 GB.}
  \label{fig:ss_user_performance_peak_rate}
\end{figure*}

\subsection{With Multiple Subscribers}
\label{sec-4-2}
Based on the results of the investigation with a single subscriber, now we study
the interaction of shaped traffic flows from multiple subscribers who share the
capacity of a common feeder link (i.e., $R_F$) in a shared access network and
their impact on user-perceived performance. The results in Sec.~\ref{sec-4-1}
suggest that increasing the token bucket size (e.g., from 1 MB to 10 MB to 100
MB for the access line rate of 100 Mbit/s) improves user-perceived performance at
the subscriber level, which is a good news from end-users' perspective. The
large token bucket size and the resulting large bursts from each subscriber's
traffic, on the other hand, may have negative impacts on the user-perceived
performance at the access level due to their interaction on the common shared
link. From ISPs' point of view, it is interesting to see how many subscribers,
given the access line rate and the TBF parameters, can be supported with
user-perceived performance non-inferior to those of a corresponding unshaped,
single-subscriber configuration. Note that it is well known that a large token
bucket size increases the deterministic end-to-end packet delay bounds of
various work-conserving scheduling disciplines with TBF-constrained traffic
\cite{zhang95:_servic}.

% Fig.\(~\)\ref{fig:ms_performance} shows representative metrics of
% user-perceived performance for multiple subscribers with various
% configurations. As expected from the results with a single subscriber, again we
% observe that the token bucket size has the greatest effect on the average
% session throughput of FTP traffic; the average session throughput of FTP traffic
% begins to deteriorate quite earlier than the average session delay of HTTP
% traffic and the DFR of UDP streaming video for all the configurations.
% \begin{figure*}[!tpb]
%   \begin{minipage}{.49\linewidth}
%     \begin{center}
%       \includegraphics*[width=.85\linewidth]{ms_dr100M_mr10M-http-dly.pdf}\\
%       \includegraphics*[width=.85\linewidth]{ms_dr100M_mr10M-ftp-thr.pdf}\\
%       \includegraphics*[width=.85\linewidth]{ms_dr100M_mr10M-video-dfr.pdf}\\
%       {\scriptsize (a)}
%     \end{center}
%   \end{minipage}
%   \hfill
%   \begin{minipage}{.49\linewidth}
%     \begin{center}
%       \includegraphics*[width=.85\linewidth]{ms_dr1G_mr90M-http-dly.pdf}\\
%       \includegraphics*[width=.85\linewidth]{ms_dr1G_mr90M-ftp-thr.pdf}\\
%       \includegraphics*[width=.85\linewidth]{ms_dr1G_mr90M-video-dfr.pdf}\\
%       {\scriptsize (b)}
%     \end{center}
%   \end{minipage}
%   \caption{User-perceived performance measures with 95 percent confidence
%     intervals for multiple subscribers with (a) access line rate of 100 Mbit/s and
%     (b) access line rate of 1 Gbit/s.}
%   \label{fig:ms_performance}
% \end{figure*}
% %%%

The results of the multivariate non-inferiority testing for shaped
configurations with respect to the reference cases (e.g., S$_{1,5}$ and
S$_{1,6}$ with $n=4$ with respect to U$_1$ with $n=4$) are summarized in
Table~\ref{tbl:ms_noninferiority}, where $max(N_{eqv})$ is defined as the
maximum number of subscribers that can be supported with user-perceived
performance non-inferior to those with the unshaped, single-subscriber
configuration with the same number of users per subscriber. The conditions for
the multivariate non-inferiority testing are the same as in Sec.~\ref{sec-4-1}.

\begin{table}[!tpb]
  \begin{center}
    \caption{Results of multivariate non-inferiority testing}
    \label{tbl:ms_noninferiority}
    \begin{threeparttable}
      \begin{tabular}{|c|c|r|r|}
        \hline
        Config. & $n$ & $max(N_{eqv})$ & $n \cdot max(N_{eqv})$\tnote{1} \\ \hline\hline\hline
        S$_{1,3}$ & 1 & 36 & 36 \\ \hline\hline
        S$_{1,4}$ & 3 & 13 & 39 \\ \hline
        \multirow{2}{*}{S$_{1,5}$} & 3 & 15 & 45 \\ \cline{2-4}
        & 4 & 11 & 44 \\ \hline
        \multirow{3}{*}{S$_{1,6}$} & 3 & 15 & 45 \\ \cline{2-4}
        & 4 & 11 & 44 \\ \cline{2-4}
        & 5 & 9 & 45 \\ \hline\hline
        S$_{1,7}$ & 7 & 6 & 42 \\ \hline
        \multirow{2}{*}{S$_{1,8}$} & 7 & 6 & 42 \\ \cline{2-4}
        & 9 & 5 & 45 \\ \hline
        \multirow{3}{*}{S$_{1,9}$} & 7 & 6 & 42 \\ \cline{2-4}
        & 9 & 5 & 45 \\ \cline{2-4}
        & 10 & 4 & 40 \\ \hline\hline\hline
        S$_{2,3}$ & 1 & 30 & 30 \\ \hline\hline
        S$_{2,4}$ & 1 & 27 & 27 \\ \hline
        S$_{2,5}$ & 1 & 27 & 27 \\ \hline
        \multirow{2}{*}{S$_{2,6}$} & 1 & 27 & 27 \\ \cline{2-4}
        & 2 & 14 & 28 \\ \hline\hline
        S$_{2,7}$ & 2 & 14 & 28 \\ \hline
        S$_{2,8}$ & 2 & 14 & 28 \\ \hline
        \multirow{2}{*}{S$_{2,9}$} & 2 & 14 & 28 \\ \cline{2-4}
        & 3 & 9 & 27 \\ \hline
      \end{tabular}
      \begin{tablenotes}
      \item[1] Total number of users in the shared access that can be supported
        with user-perceived performance non-inferior to those of unshaped,
        dedicated access.
      % \item[2] Total downstream throughput divided by access line rate.
      \end{tablenotes}
    \end{threeparttable}
  \end{center}
\end{table}

The results in Table\(~\)\ref{tbl:ms_noninferiority} show that, unlike its
impact on packet-level performance, increasing the token bucket size (given the
configurations) does not have a negative impact on the user-perceived
performance at the access level; given the average rate and the number of users
per subscriber, the configurations with larger token bucket size can support as
many subscribers as those with smaller token bucket size (e.g., S$_{1,5}$
vs. S$_{1,6}$ for $n=3,~4$) or even more (i.e., S$_{1,4}$ vs. S$_{1,5}$ and
S$_{1,6}$ for $n=3$). Considering that we can support more users per subscriber
(e.g., up to 3 users for S$_{1,4}$ vs. up to 5 users for S$_{1,6}$) with a
larger token bucket size, we can also increase the total number of users ---
i.e., $n \cdot max(N_{eqv})$ --- by proper dimensioning of TBF parameters in the
access network which can be supported with user-perceived performance
non-inferior to those with the unshaped, single-subscriber configuration with
the same number of users per subscriber.

These results suggest that the large token bucket size in ISP traffic shaping
could improve user-perceived performance of each subscriber at both subscriber
and access levels, which would be a good news not only for end-users but also
for ISPs. Our investigation in this paper, however, does not consider the
potential negative impact of the loose burst control on metro and backbone
networks, where, unlike access networks, packet-level performance measures are
still important; note that we set the backbone line rate (i.e., $R_B$) to a
value much higher than the rate of combined traffic flows from all the
subscribers in the access network to prevent it from being a bottleneck during
the experiments.

\subsubsection{With Non-Conformant Subscribers}
\label{sec-4-2-1}
We also investigated the effect of loose burst control in the presence of
non-conformant subscribers. The experiment configuration for this investigation
is shown in Fig.~\ref{fig:unbalanced_config}, where there are two groups of
subscribers, i.e., Group~1 for 10 conformant subscribers with $n=3$ and Group~2
for non-conformant ones with $n=5$. In this experiment we set $n_h$, $n_f$ and
$n_v$ to 1 for the users in Group~1, while we vary $n_f$ with $n_h$ and $n_v$
fixed to 1 for the users in Group~2 to change network load. As for a token
bucket, we consider the size of up to 10 GB, which is about a thousand times
larger than those provided by cable broadband companies through
PowerBoost/Speedboost technologies \cite{bauer11:_power}.
\begin{figure}[htb]
  \centering
  \includegraphics[angle=-90,width=.7\linewidth]{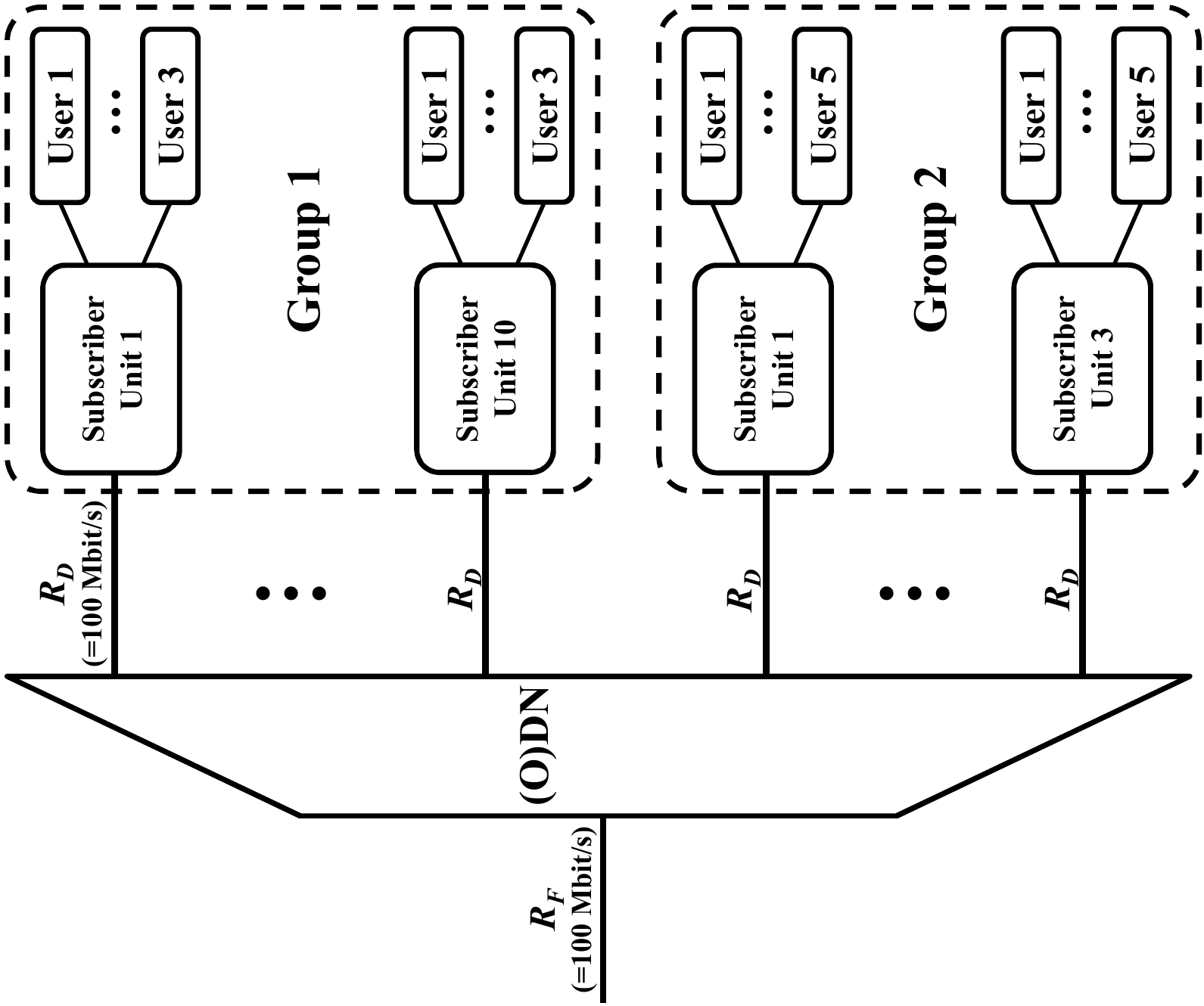}
  \caption{\label{fig:unbalanced_config}Unbalanced network configuration with
    two groups of subscribers and access line rate of 100 Mbit/s.}
\end{figure}

Fig.~\ref{fig:mu_performance} shows that the loose burst control resulting from
the large token bucket size up to 1 GB does not negatively affect user-perceived
performance with multiple subscribers even in the presence of non-conformant
subscribers; with a much larger token bucket size of 10 GB, however, the
negative effect of non-conformant subscribers on the user-perceived performance
of conformant subscribers becomes finally visible because the maximum burst
duration in this case, which is larger than average inter-session gaps of
traffic models (e.g., 30 and 180 seconds for HTTP and FTP services), makes the
impact of token bucket size and that of token generation rate virtually
indistinguishable.
%%% with RR scheduling
\begin{figure}[!tpb]
  \begin{center}
    \includegraphics*[width=.55\linewidth]{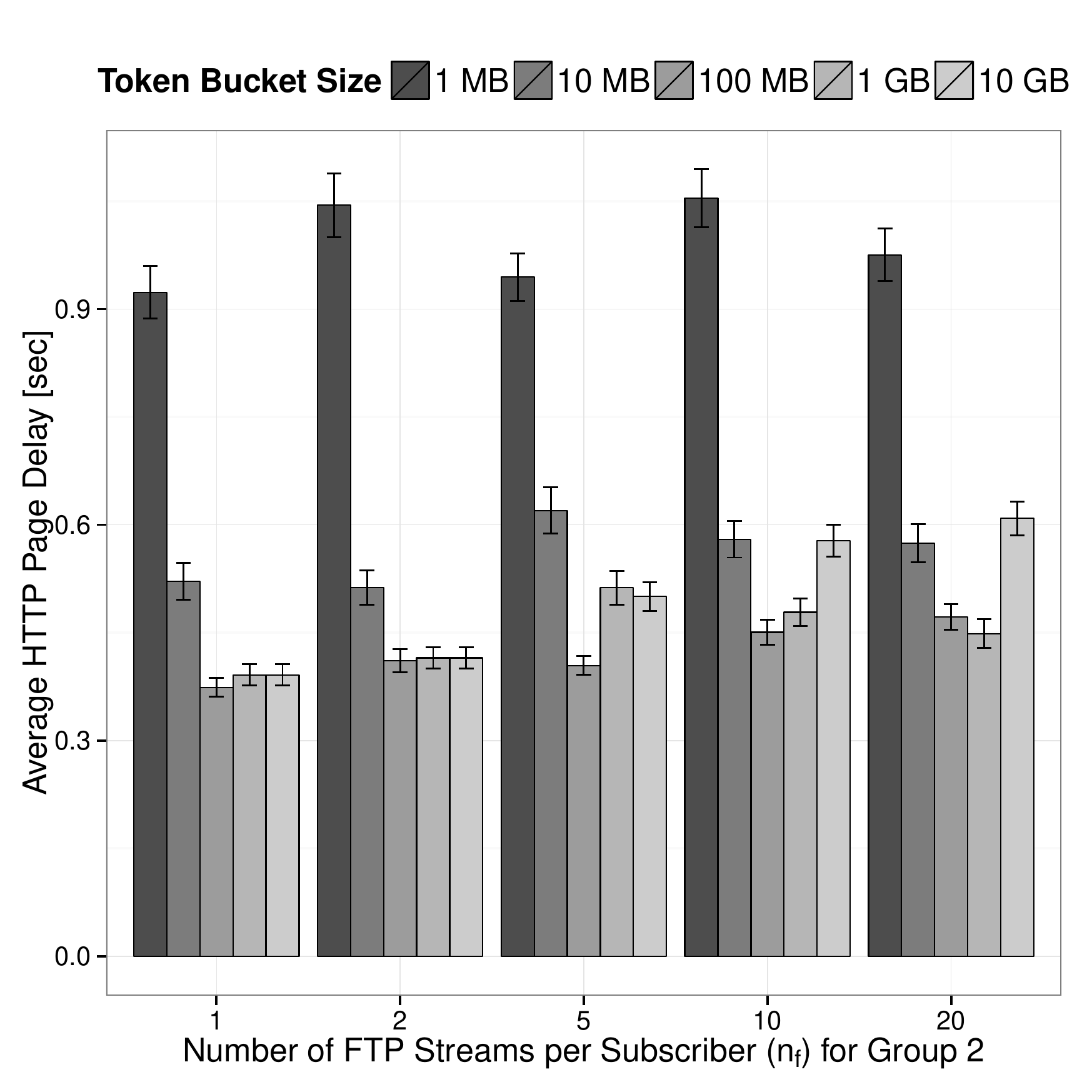}\\
    {\scriptsize (a)}
  \end{center}
  \begin{minipage}{.49\linewidth}
    \begin{center}
      \includegraphics*[width=\linewidth]{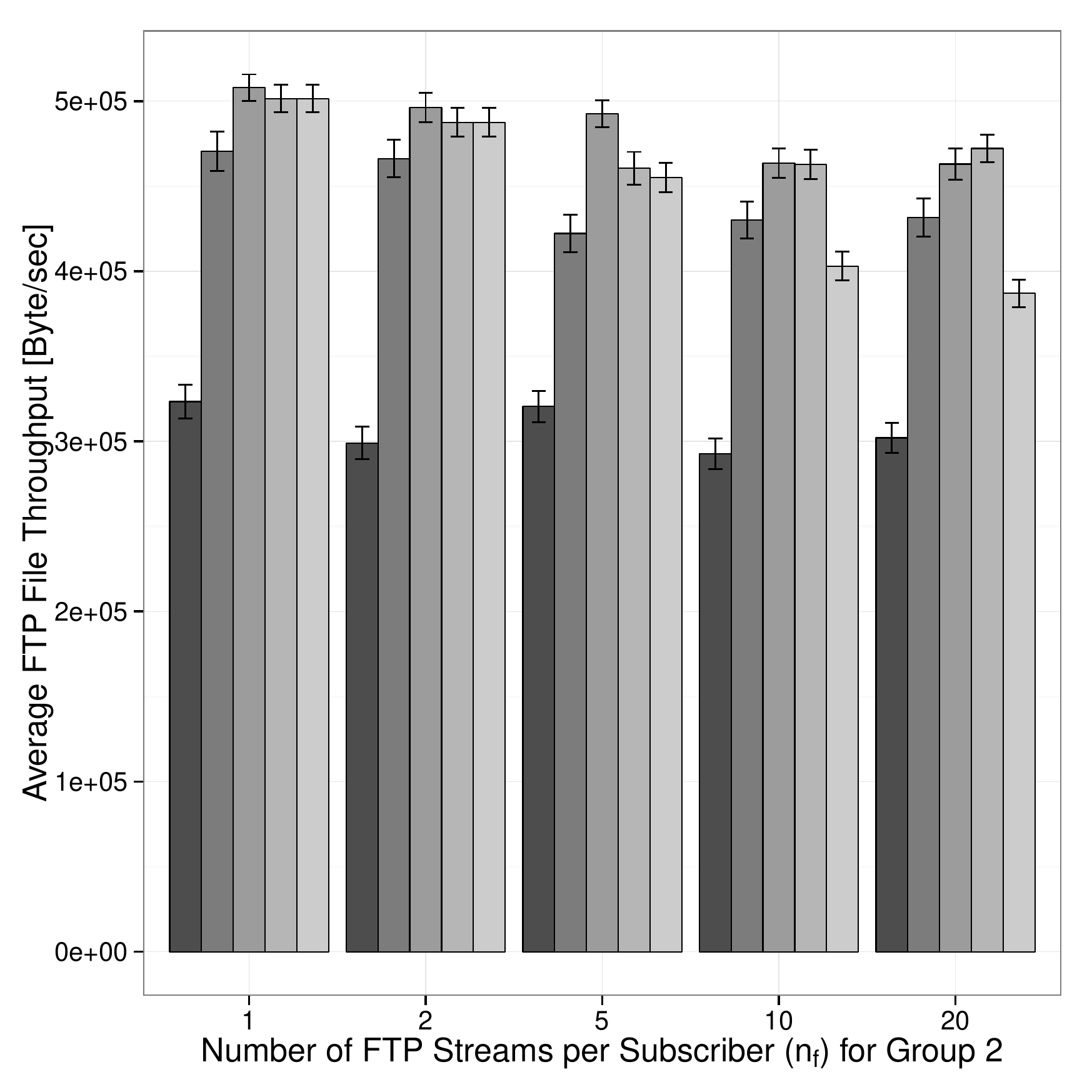}\\
      {\scriptsize (b)}
    \end{center}
  \end{minipage}
  \hfill
  \begin{minipage}{.49\linewidth}
    \begin{center}
      \includegraphics*[width=\linewidth]{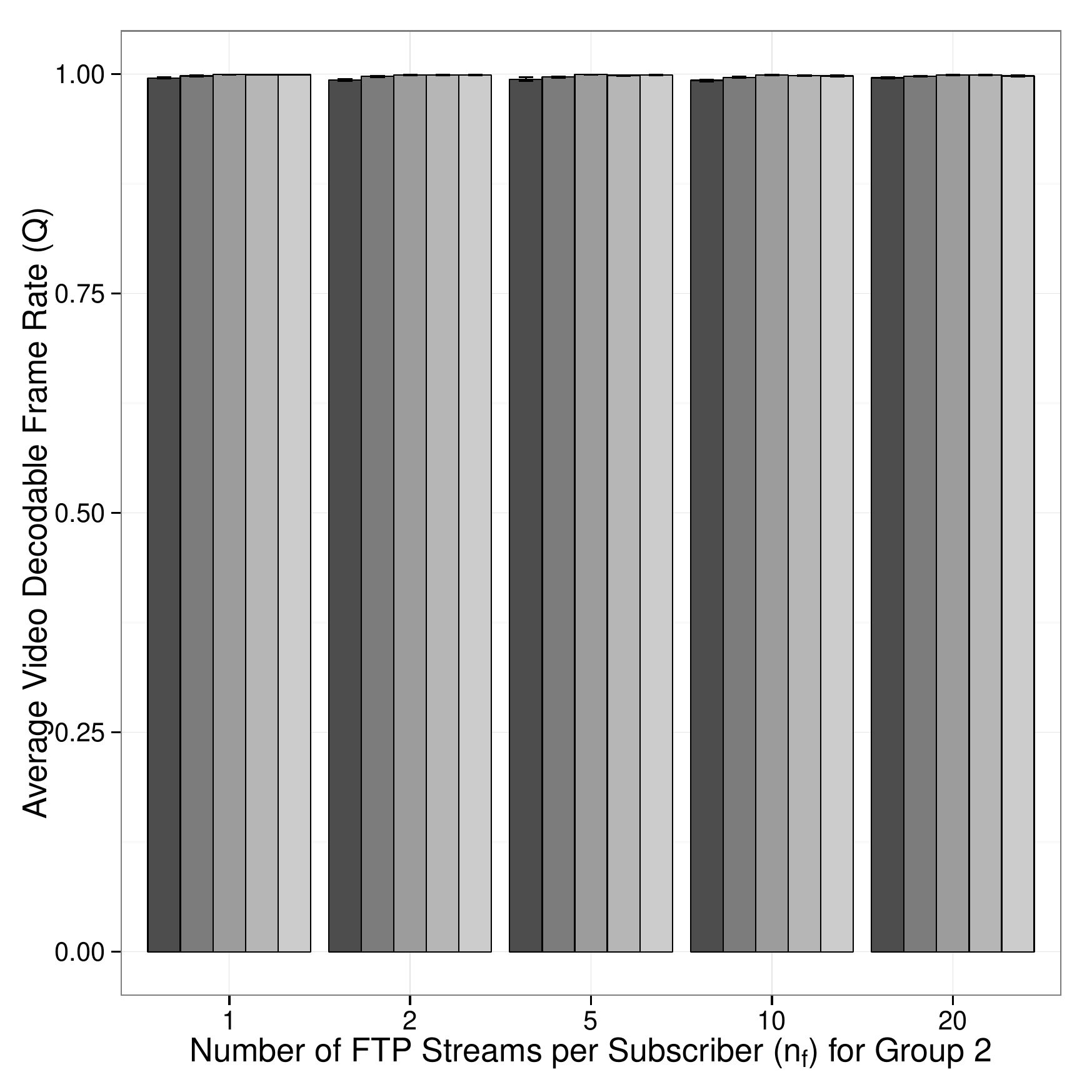}\\
      {\scriptsize (c)}
    \end{center}
  \end{minipage}
  \caption{User-perceived performance measures with 95 percent confidence
    intervals for Group~1 of the network configuration shown in
    Fig.~\ref{fig:unbalanced_config} with token generation rate of 10 Mbit/s:
    (a) Average session delay of HTTP traffic, (b) average session throughput of
    FTP traffic, and (c) decodable frame rate (Q) of UDP streaming video.}
  \label{fig:mu_performance}
\end{figure}

Fig.~\ref{fig:mu_fifo_performance} shows the results for the same configuration
of Fig.~\ref{fig:unbalanced_config} but with first-in, first-out (FIFO)
scheduling instead of round-robin, where the overall performance becomes worse
in general compared to that of round-robin scheduling. Note that, in case of
FIFO scheduling, the negative effect of non-conformant subscribers becomes clear
with the token bucket size of 1 GB.
%%% with FIFO scheduling
\begin{figure}[!tpb]
  \begin{center}
    \includegraphics*[width=.55\linewidth]{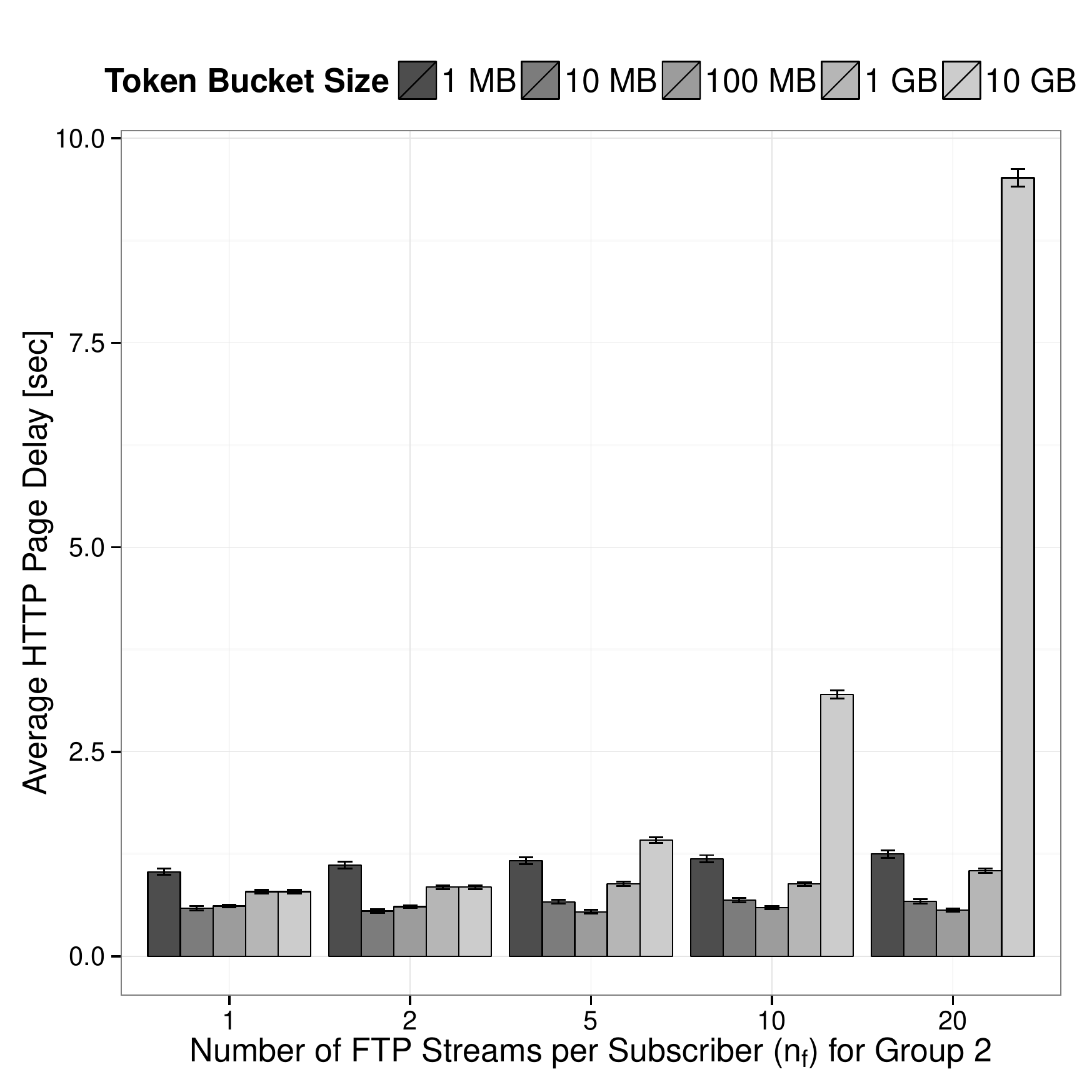}\\
    {\scriptsize (a)}
  \end{center}
  \begin{minipage}{.49\linewidth}
    \begin{center}
      \includegraphics*[width=\linewidth]{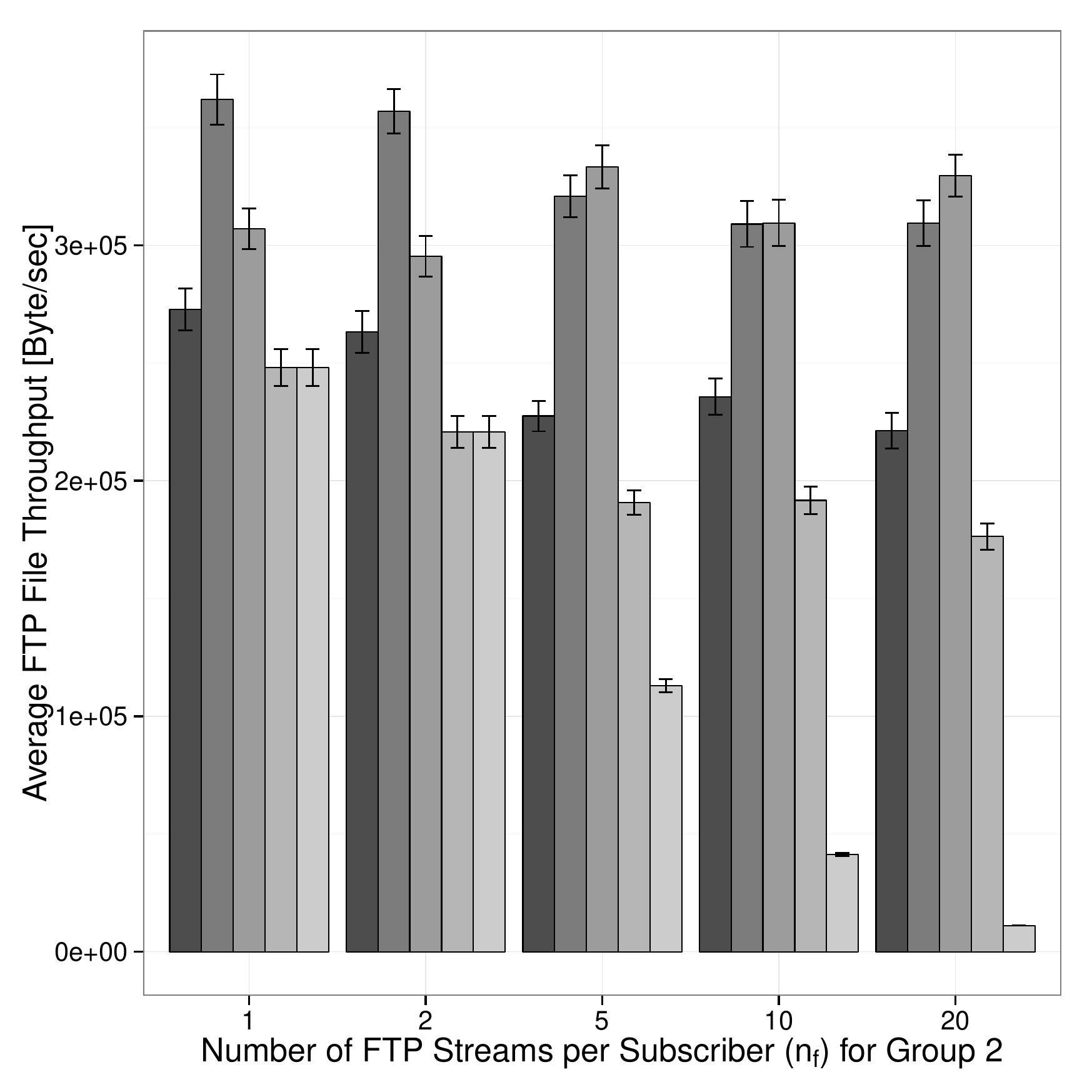}\\
      {\scriptsize (b)}
    \end{center}
  \end{minipage}
  \hfill
  \begin{minipage}{.49\linewidth}
    \begin{center}
      \includegraphics*[width=\linewidth]{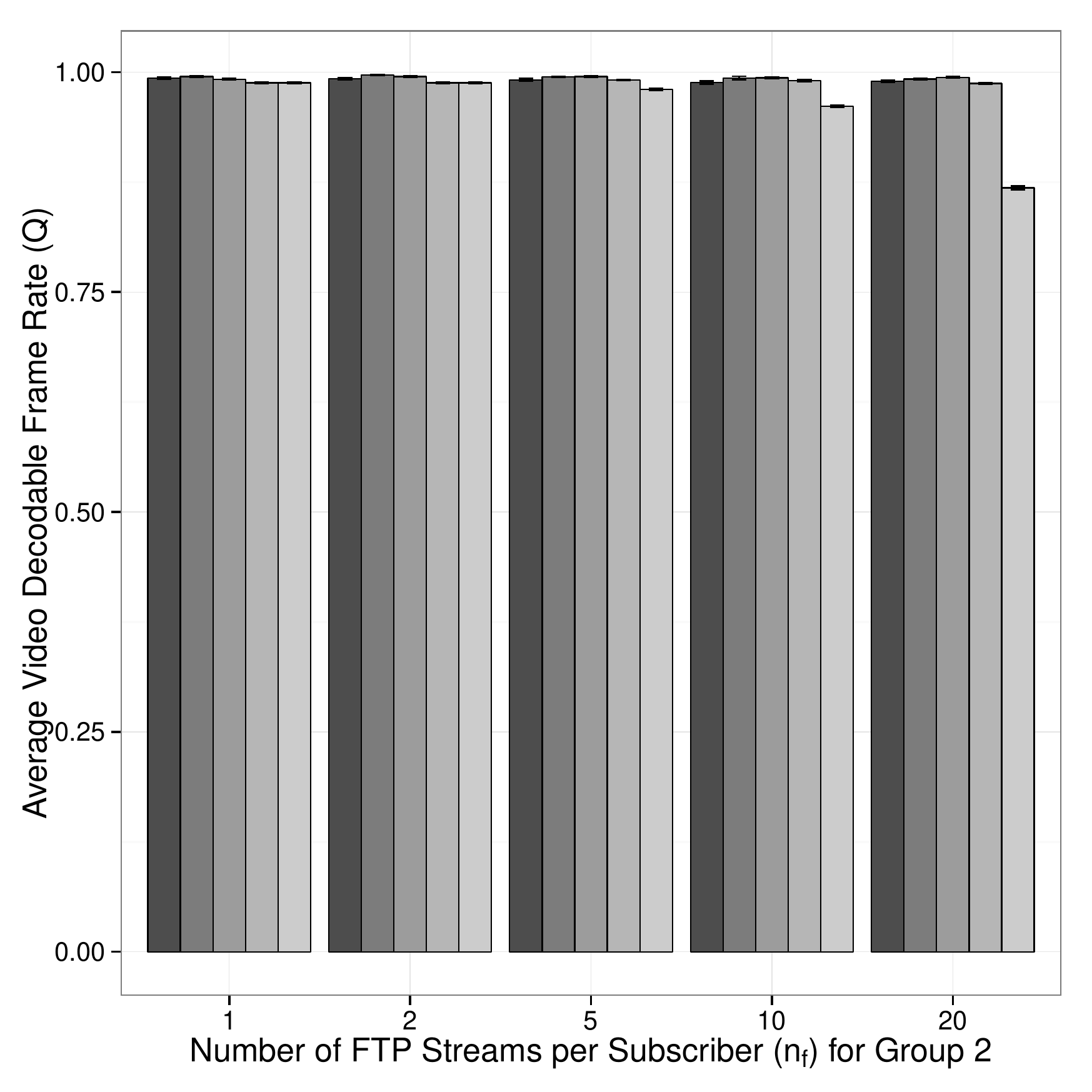}\\
      {\scriptsize (c)}
    \end{center}
  \end{minipage}
  \caption{User-perceived performance measures with 95 percent confidence
    intervals for Group~1 of the network configuration shown in
    Fig.~\ref{fig:unbalanced_config} with \textit{FIFO scheduling} and token
    generation rate of 10 Mbit/s: (a) Average session delay of HTTP traffic, (b)
    average session throughput of FTP traffic, and (c) decodable frame rate (Q)
    of UDP streaming video.}
  \label{fig:mu_fifo_performance}
\end{figure}
%%% to end the group for math mode spacing adjustment
\endgroup

\section{Conclusions}
\label{sec-5}
In this paper we have investigated the effect of ISP traffic shaping on
user-perceived performance based on user-behavior-based traffic models for HTTP
and FTP services and a real-trace-based one for streaming video and
application/session-layer performance metrics. The results from extensive
simulations show that a larger token bucket (i.e., up to 100 MB and 1 GB for
100-Mbit/s and 1-Gbit/s access line rates) provides better user-perceived
performance at both subscriber and access levels. This implies that the loose
burst control (i.e., allowing users to send their traffic at a peak rate, much
higher than the average service rate) enables to exploit well the burstiness of
real traffic in different time scales and at multiple layers --- i.e., user
behaviors (e.g., reading time in web browsing) and VBR encoding at the session
layer and TCP congestion control at the transport layer --- in statistical
multiplexing in the shared access network. Regarding any negative impact of the
loose burst control, on the other hand, we do not observe any significant
disadvantage with a larger token bucket in terms of the user-perceived
performance of conformant subscribers even in the presence of non-conformant
subscribers again up to 100 MB for the access line rate of 100 Mbit/s; with a
much larger token bucket (e.g., size of 10 GB), however, the negative effect of
non-conformant subscribers on the user-perceived performance of conformant
subscribers becomes clearly visible because the impact of token bucket size and
that of token generation rate are virtually indistinguishable in this case.

The results from the current work can provide ISPs valuable insights into the
design, deployment, and operation of the next-generation access networks from
end-users' perspective, especially for the control of peak rate and burstiness
to improve user-perceived performance for their access services. There are also
implications to researchers in the design of next-generation access
architectures and protocols, where we need to study a way to better exploit the
burstiness of end-user traffic in different time scales and at multiple layers
in statistical multiplexing. Still, we do need more investigations with a wider
range of network and traffic configurations to reach a firm conclusion on the
effect of ISP traffic shaping on user-perceived performance.

The rather different outcomes from this work, compared to those based on
traditional packet-level traffic models and performance measures, clearly show
the very importance of user-oriented research framework in the study of access
network architectures and protocols as discussed in \cite{Kim:11-2,Kim:11-4}.

% Note that the research along the lines suggested by this new framework demands
% extensive runs of realistic simulation to capture the complexity of protocols
% and the interactive nature of traffic in the study of network architectures and
% obtain test statistics for the multivariate non-inferiority testing for
% comparative analysis\footnote{The implemented simulation models, configurations,
%   and scripts for pre- and post-processing are available at
%   ``http://github.com/kyeongsoo/inet-hnrl''. }; in fact, the total number of
% simulation runs for this work is more than 4,000. The recent introduction of
% high-performance computing (HPC) clusters and the cloud computing, however,
% brings enormous computing power at a much lower cost and, in case of cloud
% computing, on-demand basis and enables researchers to run these extensive,
% large-scale network simulations in a realistic environment, which was neither
% practical nor economically feasible in the past.

One of the major difficulties in this work was the lack of
established/standardized behavioral traffic models at higher service and access
line rates. For instance, due to the lack of higher-rate FTP traffic model, we
have to use multiple FTP traffic streams to increase the load to the system
instead of single higher-rate stream for non-conformant subscribers. With
standardized sets of traffic models together with performance metrics, we could
provide both ISPs and end-users benchmarks and/or rating systems useful for
comparison shopping of broadband access services from ISPs.
% There is already a proposal called ``Internet Nutrition Labels''
% \cite{sundaresan11:_helpin_isps}, but it is still based on traditional
% network/packet-level performance measures.
In fact, all the efforts described in this paper are aiming at the creation of
new benchmarks and/or rating systems for next-generation access
% based on the proposed user-oriented research framework
and the adoption of them both by ISPs and end-users for advertising and
selecting new service plans. There is already a proposal called ``Internet
Nutrition Labels'' \cite{sundaresan11:_helpin_isps} in this regard, but it is
still based on traditional network-level performance measures. Our plan is to
have new rating systems based on the comparative analysis framework described in
Sec. 3.3 under several representative workloads, e.g., user behavior models for
web browsing, Internet voice/video calls, multimedia streaming, and online
gaming. In this way, the design, deployment, and operation of next-generation
access networks will be more energy and cost-efficient by properly managing
network resources based on actual user behaviours, not worst-case traffic, and
with a direct focus on user-perceived performance.

Another area of research for further work is the extension of the multivariate
non-inferiority testing to quantiles/percentiles, especially for delay, because
it is quite challenging to obtain not only the quantiles themselves but also
confidence intervals needed for statistical hypothesis testing
\cite{lee99:_quant,eickhoff07:_using}.

\section*{Acknowledgement}
This paper was presented in part at FOAN 2012, St. Petersburg, Russia, October
2012. This work was supported in part by Amazon Web Services (AWS) in Education
Research Grant.

% \bibliographystyle{model1-num-names}%%
% \bibliography{IEEEabrv,kks}%%

\end{document}